\documentclass[acmsmall]{acmart}
\usepackage{longtable}
\usepackage{multirow}
\usepackage{xcolor}
\definecolor{darkgreen}{rgb}{0.0, 0.5, 0.0}
\definecolor{grey}{rgb}{0.5, 0.5, 0.5}

\theoremstyle{definition}

\AtBeginDocument{%
  }

\renewcommand\footnotetextcopyrightpermission[1]{} 

\usepackage{graphicx}
\usepackage{epstopdf}
\usepackage{subcaption}
\usepackage{threeparttable}
\usepackage{wrapfig}
\usepackage{lipsum}
\usepackage[normalem]{ulem}
\usepackage{fontawesome5}
\usepackage{makecell}
\usepackage{booktabs}
\usepackage{tabularx}
\usepackage[table]{xcolor} 
\usepackage{colortbl}

\newcolumntype{L}[1]{>{\raggedright\arraybackslash}p{#1}}
\newcolumntype{C}[1]{>{\centering\arraybackslash}p{#1}}

\usepackage{hyperref}
\hypersetup{
  colorlinks=true,
  linkcolor=black,
  urlcolor=black
}

\begin{document}

\title{Large Language Models and Social Media Information Integrity: Opportunities, Challenges, and Research Directions}


\author{Junjie Xiong$^{*\dag}$}
\affiliation{%
  \institution{Missouri University of Science and Technology}
  \city{Rolla}
  \country{U.S.}}
\email{Junjiexiong@mst.edu}

\author{Zhengyuan Jiang$^*$}
\affiliation{%
  \institution{University of South Florida}
  \city{Tampa}
  \country{U.S.}}
\email{jz1@usf.edu}

\author{Xiaoran Xu$^*$}
\affiliation{%
  \institution{University of South Florida}
  \city{Tampa}
  \country{U.S.}}
\email{xiaoranxu@usf.edu}

\author{Chi Zhang$^*$}
\affiliation{%
  \institution{University of South Florida}
  \city{Tampa}
  \country{U.S.}}
\email{chiz@usf.edu}

\author{Changjia Zhu}
\affiliation{%
  \institution{University of South Florida}
  \city{Tampa}
  \country{U.S.}}
\email{changjiaz@usf.edu}

\author{Ning Wang}
\affiliation{%
  \institution{University of South Florida}
  \city{Tampa}
  \country{U.S.}}
\email{ningw@usf.edu}

\author{Mingkui Wei}
\affiliation{%
  \institution{George Mason University}
  \city{Fairfax}
  \country{U.S.}}
\email{mwei2@gmu.edu}

\author{Zhuo Lu}
\affiliation{%
  \institution{University of South Florida}
  \city{Tampa}
  \country{U.S.}}
\email{zhuolu@usf.edu}

\author{Yao Liu}
\affiliation{%
  \institution{University of South Florida}
  \city{Tampa}
  \country{U.S.}}
\email{yliu21@usf.edu}

\author{Lingyao Li $^\dag$}
\affiliation{%
  \institution{University of South Florida}
  \city{Tampa}
  \country{U.S.}}
\email{lingyaol@usf.edu}

\thanks{$^\dag$Corresponding author.}

\renewcommand{\shortauthors}{Xiong et al.}



\begin{abstract}
    Large Language Models (LLMs) have emerged as powerful tools that impact information integrity on social media platforms. This comprehensive review examines the dual role of LLMs in both facilitating and mitigating various information integrity challenges, including misinformation, disinformation, fake news, social bots, and privacy concerns. \textcolor{black}{We conduct a comprehensive review of the literature from 2019 to 2024, screening 1048 studies and performing an in-depth analysis of 215 representative papers. This systematic approach allows us to identify key patterns in how LLMs influence the information security in social media ecosystems.} Through a systematic analysis of papers from multiple databases, our findings reveal that while LLMs can enhance detection capabilities for malicious content and enable sophisticated defense mechanisms, they simultaneously pose risks by enabling the generation of highly convincing, deceptive content. We categorize and analyze the potential and challenges across different dimensions of information integrity, examining technical capabilities, ethical implications, and privacy concerns. The study demonstrates critical gaps in current approaches, particularly in cross-lingual detection, real-time monitoring, and privacy-preserving implementations. We conclude by proposing future research directions and recommendations for stakeholders to leverage LLMs while mitigating risks in social media information integrity. 
\end{abstract}


\newpage
\maketitle

\section{Introduction}
Social media platforms have become indispensable tools for modern communication, connecting users around the world and facilitating real-time information sharing. However, these digital platforms remain vulnerable to critical information security threats, including misinformation, disinformation, and the propagation of biased or toxic content \cite{grinberg2019fake,tokita2024measuring}. These vulnerabilities not only erode public trust but also present substantial challenges in maintaining the integrity of digital systems. The growing risks associated with the dissemination of false information and cyber threats underscore the urgent need to safeguard online environments \cite{Tsvetkova2024-cz}.

Large Language Models (LLMs) have rapidly become embedded in today's digital platforms. These AI systems, exemplified by GPT-style models, drive a range of applications from conversational agents to intelligent digital assistants and content creation tools. Social media platforms increasingly leverage LLM-driven bots to engage users or automate customer service, while writing assistants use them to generate articles, marketing copy, and code~\citet{zhou2024correcting, lu2024corporate}. The integration of LLMs is reshaping how information is produced and consumed in digital systems. On the positive side, LLMs have shown potential in detecting and mitigating harmful content \cite{lee2024train, Yang2024-tq}. LLM-powered applications can also help moderate content and flag policy violations at scale, complementing human moderators in keeping online communities civil~\cite{Feng2024-gf, Haynes2021-ga}. These applications show that LLMs carry significant implications for the quality of public discourse and the security of digital environments. 

Despite their capabilities, the integration of LLMs can also impact information integrity. A major concern is their tendency to ``hallucinate,'' generating text that appears authentic and authoritative but may be false or misleading \cite{Shah2024-js,Christensen2024-qq}. For example, recent studies show that LLMs can produce election-related misinformation that is almost indistinguishable from human-written content \cite{Borji2023-hb,spitale2023ai}. Additionally, they have been known to generate misinformation or fake news, including hateful or offensive language, particularly when trained on toxic datasets or manipulated through jailbreak attacks~\cite{Haynes2021-ga, Lin2024-xa}. High-profile incidents, such as an AI chatbot coerced into spreading racist and offensive messages, illustrate this risk. Moreover, because LLMs learn from vast internet datasets, they can internalize and reinforce societal biases, associating certain groups with negative stereotypes and potentially harming marginalized communities \cite{Lin2024-xa, Heaton2024-aw, Duncan2023-xk}. When combined with social media's amplification mechanisms, LLM-generated misinformation or harmful content can spread rapidly, misleading users on critical topics and further complicating efforts to maintain the integrity of digital discourse.

\begin{figure}[b]
    \centering
    \includegraphics[width=0.8\linewidth]{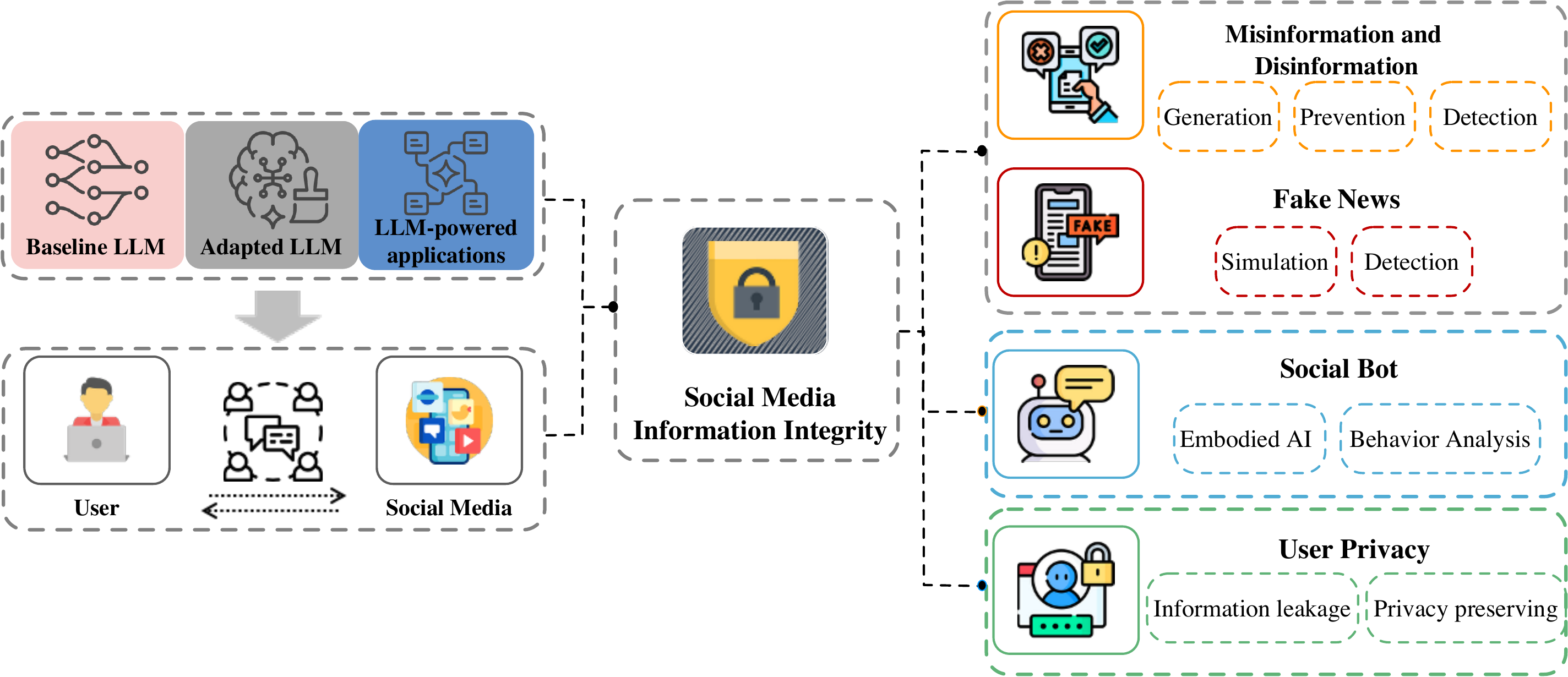}
    \caption{{The Integration and Impact of LLMs in Social Media Information Integrity. 
    }
    }
    \label{fig: Threat model story (waiting for update)}
\end{figure}

To mitigate these issues, researchers and practitioners have proposed a variety of mitigation strategies to align LLM behavior with ethical and safety standards \cite{Heaton2024-aw,huang-etal-2024-flames}. One typical approach is the reinforcement learning from human feedback (RLHF) alignment pipeline, which fine-tunes models using human input to reinforce desirable behaviors~\citet{bai2022training, kaufmann2024survey}. By training on examples of preferred behaviors and using human judgments to reward desirable answers, LLMs can be steered toward more helpful and harmless responses. In addition, researchers have developed data curation pipelines to filter out misinformation from training corpora \cite{staab2023beyond}. Such dataset-level interventions can preempt many issues before the model engages with users. Another typical strategy is the adversarial testing \cite{Feng2024-gf}. For example, adversarial prompts are designed to find weaknesses. By learning from these adversarial cases (e.g., either via additional fine-tuning or adjusting the model's safety filters), developers can strengthen the model against malicious exploitation. These typical strategies can help ensure that known failure modes are addressed before online users encounter them.

\textcolor{black}{Even though recent research into LLM security, bias mitigation, and ethical AI has grown exponentially, existing literature surveys often prioritize static assessments of general model capabilities, leaving a significant void regarding the dynamic and interconnected risks within social media ecosystems. Specifically, leading surveys on adversarial security systematically categorize technical vulnerabilities like prompt injection \cite{das2025security} but overlook the ``asymmetric co-evolutionary dynamics'' inherent to social media environments \cite{dukic2020you}. For example, the attackers can iterate content mutations at low costs while defenders operate under high verification burdens and user-trust constraints. Similarly, comprehensive reviews of bias and fairness offer robust taxonomies for evaluation metrics \cite{gallegos2024bias} but primarily focus on ``point-in-time'' analyses, failing to capture how biases evolve in real-time alongside shifting social trends and user interactions \cite{barbulescu2024each}. These works also exhibit significant gaps in cross-lingual and cross-cultural robustness, often centering on Anglocentric contexts \cite{duncan2023unmasking} while neglecting the specific harms faced by marginalized communities in diverse linguistic settings \cite{ahmed2023score_isall_you_need}. Furthermore, safety evaluation frameworks like ``The Scales of Justitia'' \cite{liu2025scales} typically focus on whether models follow instructions, yet they fail to address systemic ``responsibility gaps'' \cite{santoni2021four}---the ambiguous allocation of moral and legal accountability between developers, platform operators, and users, or the potential for second-order harms such as the ``chilling effect'' of automated moderation\cite{bonechi2024development}. Our study addresses these systematic gaps by establishing a multi-dimensional framework that integrates the content layer (information disorder), agent layer (social bots), and infrastructure layer (privacy and ethics), redefining information integrity as a comprehensive ``evidence-building'' process designed to counter multimodal attacks \cite{shan2024multimodal} and resolve complex multi-stakeholder responsibility chains, motivating the following questions:}

\begin{itemize}
    \item \textbf{RQ1.} What is the potential of LLMs in enhancing social media information integrity?
    \item \textbf{RQ2.} How do LLMs challenge the detection and mitigation of information integrity issues?
    \item \textbf{RQ3.} To what extent do ethical and security implications emerge from LLMs' role in information integrity?
\end{itemize}

In addressing these research questions, we propose a multi-dimensional framework for social media information integrity in the LLM era, systematically analyze LLM-driven potentials and risks, and outline future research and governance directions.

\section{Backgrounds}

\subsection{Concepts and Threat Model}
\textcolor{black}{In this survey, we treat \emph{information integrity} as a multi-dimensional target rather than a self-evident synonym for factual accuracy. In social-media environments, integrity encompasses (i) \emph{veracity}, namely whether claims are factually correct, (ii) \emph{grounding and sourcing}, meaning that information is supported by traceable and credible evidence, (iii) \emph{authenticity of actors and context}, including the absence of coordinated manipulation, impersonation, or synthetic amplification, and (iv) \emph{reliability under adversarial framing}, capturing misleading presentation or selective distortion even when content is partially true. Correspondingly, integrity can be operationalized at a systemic level through measurable indicators such as exposure-weighted accuracy, calibration and uncertainty awareness, provenance and attribution quality, rates of coordinated inauthentic behavior, and downstream impacts on public trust and belief formation. This explicit operational framing clarifies the target variable for evaluating both threats and mitigation strategies throughout the survey. \emph{Information security} refers to protecting digital assets and systems under the confidentiality, integrity, and availability, including identity, access control, and system abuse. \emph{Safety} refers to preventing harmful outcomes from model behavior or deployment, including misuse and unintended impacts. These concepts overlap but are not interchangeable: we focus on information integrity as the primary construct while treating security and safety as enabling conditions and adjacent risk domains. We study LLM-enabled threats in which attackers use automation to generate, personalize, and iteratively mutate deceptive content while coordinating large-scale dissemination and probing moderation systems to evade detection. Actors may range from operators to scammers or ideologically motivated groups. Defenders operate under resource and latency constraints and face asymmetric error costs: false negatives allow harms to scale, whereas false positives erode trust. We evaluate integrity using measurable targets such as reduced exposure to deceptive content, campaign disruption, and privacy protection. Users act as both amplifiers of harm and contributors to resilience, underscoring the need for both technical and user interventions.}

\subsection{What are the information security issues in social media?}


\textcolor{black}{The challenge of false information has become increasingly critical as social media platforms emerge as primary news sources, with reports indicating that 54\% of US adults now obtain their news through platforms like Facebook and YouTube~\cite{SocialNewsFact}. Our analysis of social media security issues reveals three interconnected categories that represent the primary challenges in maintaining information integrity. First, information disorder—encompassing misinformation, disinformation, and fake news—represents the content-level challenges. This category includes both unintentional spread of incorrect information and deliberate manipulation of facts, with studies showing that pseudoscience, conspiracy theories, lies, and deepfakes constitute significant threats~\cite{altay2023survey}. Second, social bots represent the agent-level challenges, serving as primary vectors for amplifying and automating the spread of information integrity issues. These automated systems, enhanced by LLMs, can now generate highly convincing content while mimicking human behavior patterns~\cite{staab2023beyond}. Third, user privacy concerns represent the infrastructure-level challenges, where the need to protect personal information intersects with the requirements for effective content moderation and bot detection~\cite{nissenbaum2019contextual}. The advent of LLMs has significantly amplified these challenges across all three categories. For example, models like GPT-3 can generate highly convincing false information that humans find both comprehensible and compelling ~\cite{spitale2023ai}. This capability becomes particularly concerning in critical domains such as health-related topics~\cite{sharevski2023talking,brandt2023ai}, even in the presence of platform safeguards such as TikTok's misinformation guard. We provide the complete definitions of these social media information integrity concepts in the Table~\ref{tab:hot-definitions}. We also note that although we structure the survey around three primary domains, this taxonomy serves as a high-level organizational framework rather than a rigid or mutually exclusive partition, as cross-cutting dimensions such as multimodal analysis, cross-lingual evaluation, and low-resource settings span multiple categories and are therefore discussed within relevant sections.}

Information Disorder, defined in \citet{wardle2017information, shu2020mining}, includes false information intended to harm and factual information being used to manipulate. In the context of social media, we focus on the harmful aspects, known as misinformation, disinformation, and fake news. LLM can further complicate the information ecosystem. On one hand, LLMs can be exploited to automatically generate misleading content, impersonate users, or flood platforms with synthetic narratives, thereby intensifying the spread of information disorder. On the other hand, these models also offer potential countermeasures: they can be fine-tuned to detect falsehoods, identify manipulation patterns, and assist in fact-checking at scale.
Misinformation and Disinformation represent two distinct but related threats to information integrity. Misinformation refers to false information shared without malicious intent ~\cite{wardle2017information}, while disinformation involves deliberately created false content intended to cause harm~\cite{wardle2017information}. The United Nations High Commissioner for Refugees (UNHCR) further distinguishes these two concepts by emphasizing that disinformation includes specific malicious content types such as hoaxes and propaganda~\cite{UNHCR}. 

\begin{table}[htbp]
    \small
  \centering
    \caption{{Definitions and Sources of Key Information Integrity Concepts in Social Media.}}

  \begin{tabular}{l | p{3.5cm} | p{5.2cm} | p{1.5cm}}
  \toprule
  \textbf{Category} & \textbf{Author and Year} & \textbf{Definition} & \textbf{Origin} \\
    \midrule
  \textbf{Misinformation} 
  & Altay et al. (2023)\cite{altay2023survey} & False and misleading information. & Academia\\
  \cmidrule(lr){2-4}
  &  Wardle et al.(2017)\cite{wardle2017information} & Information that is false but not created with the intention of causing harm. & Academia \\
  \cmidrule(lr){2-4}
  & UNHCR (2021)\cite{UNHCR} &  Misinformation is false or inaccurate information. Examples include rumors, insults, and pranks. & NGO\\

  \midrule
  
  \textbf{Disinformation} &  Wardle et al.(2017)\cite{wardle2017information} & Information that is false and is knowingly shared to cause harm. & Academia \\
  \cmidrule(lr){2-4}
  & UNHCR (2021)\cite{UNHCR} &  Disinformation is deliberate and includes malicious content such as hoaxes, spear phishing, and propaganda. It spreads fear and suspicion among the population.& NGO\\

  \midrule
  
  \textbf{Fake News} &  Cooke et al.(2017) \cite{cooke2017posttruth} & False and often sensational information disseminated under the guise of news reporting, yet the term has evolved over time and has become synonymous with the spread of false information  & Academic \\
   \cmidrule(lr){2-4}
   & Allcott et al. (2017) \cite{allcott2017social} & News articles that are intentionally and verifiably false and could mislead readers. & Academic \\
  
  \midrule
  \textbf{Social Bot} &  Staab et al.(2023)~\cite{staab2023beyond}\newline Lyu et al.(2023)~\cite{lyu2023gpt} & Social bots capable of inferring and utilizing multimodal capabilities, effectively processing and generating both textual and visual content.  & Academic \\
  \cmidrule(lr){2-4}
   & Cloudflare (2025)~\cite{cloudflare_social_bots} & Social bots are automated programs that mimic human users, operating partially or fully on their own, with many being used for deceptive or harmful purposes. & Industry \\
  \midrule
  \textbf{User Privacy} &  Helen et al.(2019)~\cite{nissenbaum2019contextual}
   & The right of users to control information flows and disclosures about themselves according to contextual norms and social expectations. & Academia \\
  \cmidrule(lr){2-4}
  & Facebook (2023)~\cite{facebook2023privacy} & Providing users transparency and practical control over how their personal data is collected, used, and shared, consistent with regulatory compliance. & Industry \\
  \bottomrule
  \end{tabular}

  \label{tab:hot-definitions}
  \end{table}

Fake News has overlaps with misinformation and disinformation, but is distinguished through three key characteristics ~\cite{cooke2017posttruth, allcott2017social}: its deliberate mimicry of legitimate news media formatting and style; its exploitation of established news distribution channels; and its intent to deceive by leveraging the credibility traditionally associated with journalism. Fake news refers to objectively false information presented in various forms, including news articles, public statements, speeches, and social media posts~\cite{zhou2020survey}. Its core feature lies in the deviation from objective facts, regardless of the publisher's motivation~\cite{vosoughi2018spread}. Whether created for malicious deception, entertainment, satire, or through unintentional dissemination, any content that contradicts factual truth falls under this category. This phenomenon has a long history of influencing political views and social discourse~\cite{lazer2018science}.

Social bots are automated programs that mimic human interaction on social media platforms, and their capabilities have significantly advanced with the integration of LLMs. While traditional bots relied on predefined scripts and rule-based automation \cite{Ghanem2020-kg, Gupta2019-vf}, LLM-enhanced bots leverage large-scale language models to generate highly contextual, personalized, and human-like content \cite{Li2023-et, Yang2024-tq}. These systems demonstrate sophisticated camouflage abilities, adaptive behavior through interaction learning \cite{Feng2024-gf}, and multimodal processing of text and images \cite{staab2023beyond, lyu2023gpt}. They can infer sensitive personal attributes with high accuracy and craft convincing phishing or social engineering content \cite{Collier2024-ng}, substantially increasing the scale and subtlety of automated manipulation. At the same time, these advancements demonstrate opportunities to employ LLMs in more adaptive and context-aware bot detection and moderation frameworks \cite{Ghanem2023-ku}.

User privacy encompasses individuals’ rights to control how their information flows within specific social contexts \cite{nissenbaum2019contextual}, extending beyond industry-focused notions of transparency and regulatory compliance \cite{facebook2023privacy}. The integration of LLMs into social media platforms introduces heightened privacy risks through powerful inference and memorization capabilities. Studies show that LLMs can deduce sensitive attributes from pseudonymized content and reconstruct detailed user profiles from fragmented data \cite{staab2023beyond}, while memorization and extraction attacks expose risks of training data leakage \cite{kandpal2023memorization, duan2025latent}. Cross-platform linkage tools further amplify these risks by automatically connecting user identities across services \cite{treves2023rurlman}, and model inversion or membership inference attacks can reveal participation in training datasets \cite{carlini2021extracting, zhang2024mia}. These vulnerabilities become particularly concerning in real-time social media environments, where behavioral traces may unintentionally disclose personal habits, health status, or relationships.

\color{black}



\subsection{Why does LLM raise information security issues in social media?}

The rapid evolution of social media platforms has fundamentally transformed how information is created, shared, and consumed online \cite{grinberg2019fake, tokita2024measuring}. Within this landscape, LLMs have emerged as transformative technologies that significantly impact information security dynamics \cite{staab2023beyond, supriya2022p}. Their deep integration into social media systems, from content moderation to user interaction, has created new opportunities and challenges that demand careful examination \cite{dogan2023catch, carlini2021extracting}. This involvement stems from several fundamental factors:



\textbf{Technical Capabilities and Limitations:} LLMs' advanced natural language processing capabilities enable them to generate highly convincing content that can be indistinguishable from human-written text \cite{Yang2024-tq}, while exhibiting tendencies to hallucinate \cite{Shah2024-js}. \textcolor{black}{This dual-use characteristic is reflected in prior studies that explore LLMs both as detection tools and as enablers of more evasive attacks. Under specific benchmarks and controlled experimental settings, some works report high detection accuracy (e.g., exceeding 98\%) \cite{Ghanem2023-ku}, while others document non-trivial evasion rates for LLM-enhanced bots (e.g., up to 29.6\%) \cite{Feng2024-gf}.}
\textcolor{black}{LLMs exhibit a distinctive dual-use profile: the same generative and reasoning capabilities that enable assistance, moderation, and detection can also be repurposed for scalable manipulation, evasion, and privacy-invasive inference. Rather than a symmetric ``technological arms race'', the attacker and defender interaction is more accurately characterized as an \emph{asymmetric co-evolutionary system}. Offense can often iterate at low marginal cost through automated content mutation, rapid probing of detection boundaries, and reuse of attack templates across platforms, while defense must absorb higher verification costs, manage false positives that erode user trust, and operate under latency and computational constraints. These asymmetries induce selective pressures: attackers are incentivized to maximize reach and evasion under limited risk, whereas defenders must balance harm reduction against robustness, transparency, and user experience.}


\textbf{Scale and Privacy Implications:} The integration of LLMs enables automated content generation at unprecedented scales \cite{Yang2024-tq}, while raising substantial privacy concerns through their sophisticated inference capabilities. These models can extract and correlate personal information from seemingly unrelated data points, potentially revealing sensitive attributes such as location, demographics, and behavioral patterns \cite{staab2023beyond}. The risk is further amplified by LLMs' ability to process and analyze vast amounts of historical user data, where patterns in content creation and interaction histories may inadvertently expose private information. This creates fundamental tensions between leveraging LLMs' powerful functionalities for content generation and ensuring robust privacy protection \cite{Puppala2024-mx}.
These characteristics make LLMs central to both the problems and solutions in social media information security. Understanding these fundamental factors is crucial for developing effective strategies to harness LLMs' potential while mitigating their risks. In the following sections, we present a comprehensive scope review of both the opportunities and challenges that emerge from LLMs' integration into social media information security systems.



In this paper, we identify three critical areas that warrant in-depth analysis in social media security: information disorder, social bots, and privacy concerns. The first area examines misinformation and disinformation detection using enhanced LLM capabilities and automated systems. The second area focuses on LLM-enhanced detection techniques for identifying automated accounts and coordinated networks. The third area addresses privacy-preserving mechanisms while ensuring information integrity. These interconnected areas represent the key challenges where LLMs present both risks and opportunities in social media environments.

\begin{wrapfigure}{r}{0.4\linewidth}
    \vspace{-35pt}   
    \centering
    \includegraphics[width=1\linewidth]{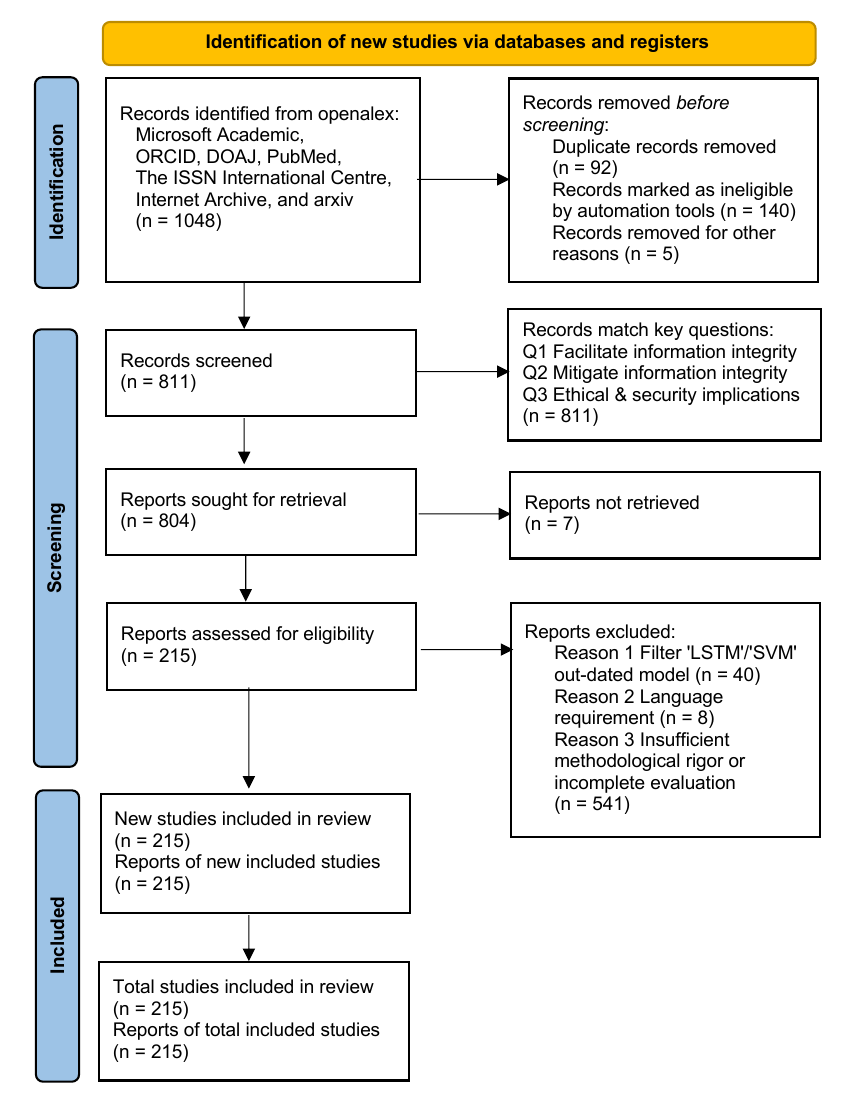}
    \caption{\textcolor{black}{PRISMA flow of the study screening.}}
    \label{LLM_Flow2}
\end{wrapfigure}

\section{Data, Methods and Initial Findings}

\subsection{Data preparation}

\textcolor{black}{
Records are retrieved via OpenAlex \cite{openalexsource2024}, which aggregates scholarly metadata primarily from Microsoft Academic Graph/Open Academic Graph and Crossref, with additional contributions from ORCID, DOAJ, PubMed, the ISSN International Centre, the Internet Archive, and arXiv.
Given the rapid pace of LLM and information integrity research, we include a limited number of arXiv preprints to capture emerging results that have not yet appeared in archival venues. We explicitly distinguish preprint-only studies from peer-reviewed conference and journal publications and treat their findings as preliminary. We double-check cited preprints for updated peer-reviewed versions and revised 14 references accordingly.}

To identify relevant literature, we query OpenAlex using keywords from three core categories: ``social media,'' ``online platform,'' and ``large language model.'' The full data preparation and filtering process is illustrated in Figure \ref{LLM_Flow2}. We retrieve 1,048 papers matching our keyword criteria during this process. After removing duplicates—both identical titles and redundant formats (e.g., preprint and published versions of the same work)—we retain 956 unique records. Next, we apply a multi-stage screening process to ensure the remaining records align with the focus of our study. This involved three key filters: (1) \textbf{Topic Matching}: we retain papers addressing our core themes, including how LLMs facilitate information integrity, mitigate information integrity challenges, and ethical or security implications on social media field; (2) \textbf{Outdated Model Filtering}: \textcolor{black}{is applied to exclude studies in which traditional machine learning or early deep learning architectures (e.g., SVM, standalone CNNs, or standalone LSTMs) served as the primary detection framework. Hybrid approaches that incorporated LLMs or Transformer-based models as the core component, while using auxiliary modules such as LSTMs for sequence aggregation or feature refinement, were retained, as these works fall within the scope of LLM-centered research; }(3) \textbf{Paper Quality Verification}: we select papers that met our criteria of novelty and technical impact in addressing capabilities or challenges of LLMs in social media. 

\begin{wrapfigure}{r}{0.4\linewidth}
    \vspace{-40pt}   
\centering
\begin{minipage}{1\linewidth}
\centering
\includegraphics[width=\linewidth]{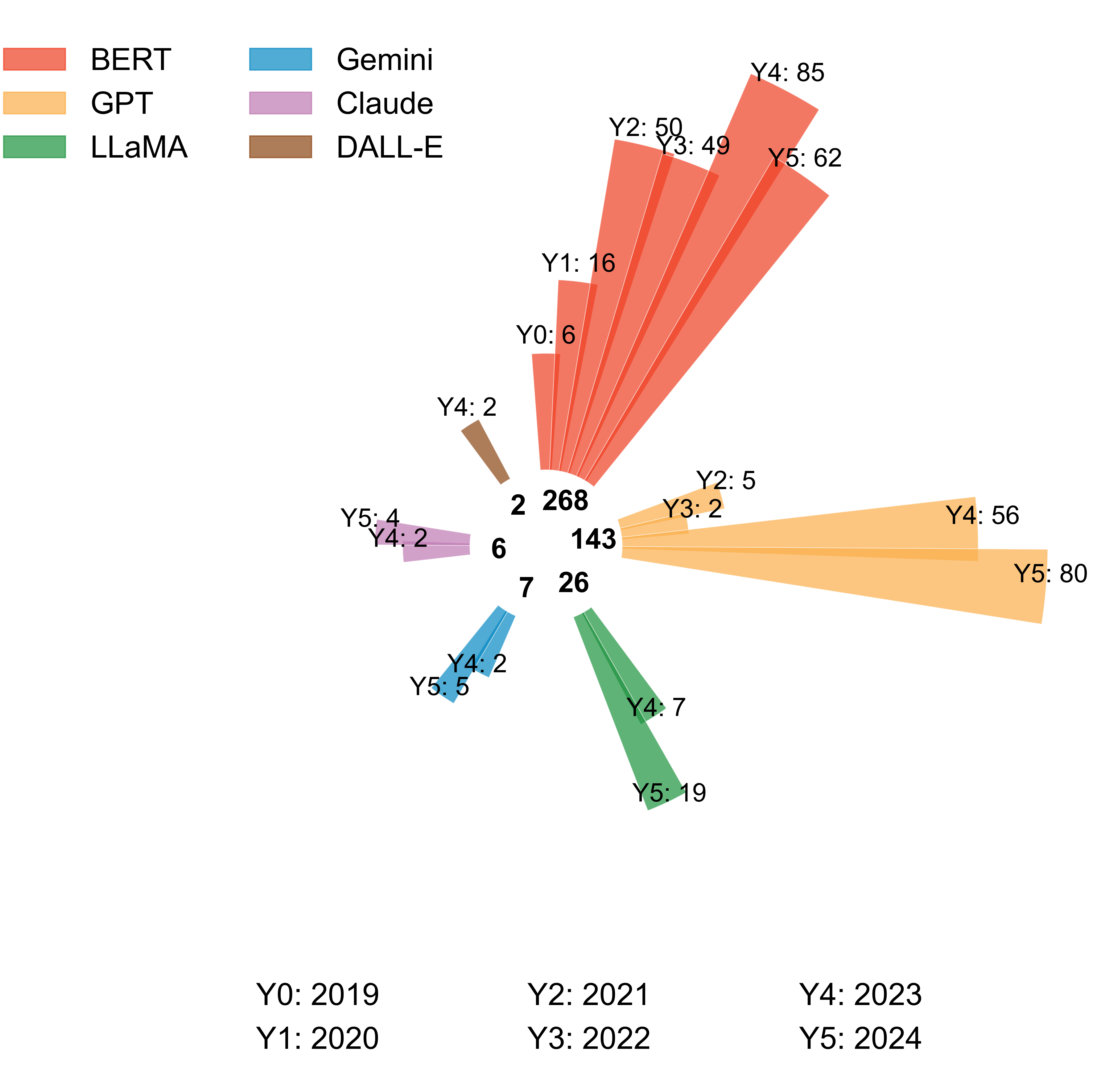}
\caption{Model popularity. Visualization of the yearly prevalence of different LLMs.}
\label{fig:data-processing1}
\end{minipage}
\vspace{-12pt}
\end{wrapfigure}

\subsection{Data Description}


To comprehensively understand the current landscape of LLMs in social media integrity research, our analysis focuses on four key dimensions: model adoption (types and scales of LLMs being utilized), architectural evolution (development in model structures and capabilities), research focus areas (primary integrity challenges being addressed), and platform influence (impact across different social media platforms). These dimensions are selected to capture both the technical advancement of LLM applications and practical impacts. 

\textbf{Model Adoption and Evolution:} The model popularity analysis (Figure \ref{fig:data-processing1}) \textcolor{black}{indicates that a large proportion of the sampled studies employ GPT-series models in social media integrity research. This preference may be attributed to several factors: GPT models' accessibility through well-documented APIs, their extensive pre-training on diverse datasets, and their demonstrated capabilities in language understanding tasks. The growing adoption of alternative models like LLaMA and Bloom might reflect researchers' interest in open-source alternatives and specialized architectures that can be fine-tuned for specific integrity tasks. This diversification in model suggests an evolving research landscape where different models serve complementary roles in addressing various aspects of information integrity.}

\begin{wrapfigure}{r}{0.4\linewidth}
\vspace{-40pt}
\centering
\begin{minipage}{1\linewidth}
\centering
\includegraphics[width=\linewidth]{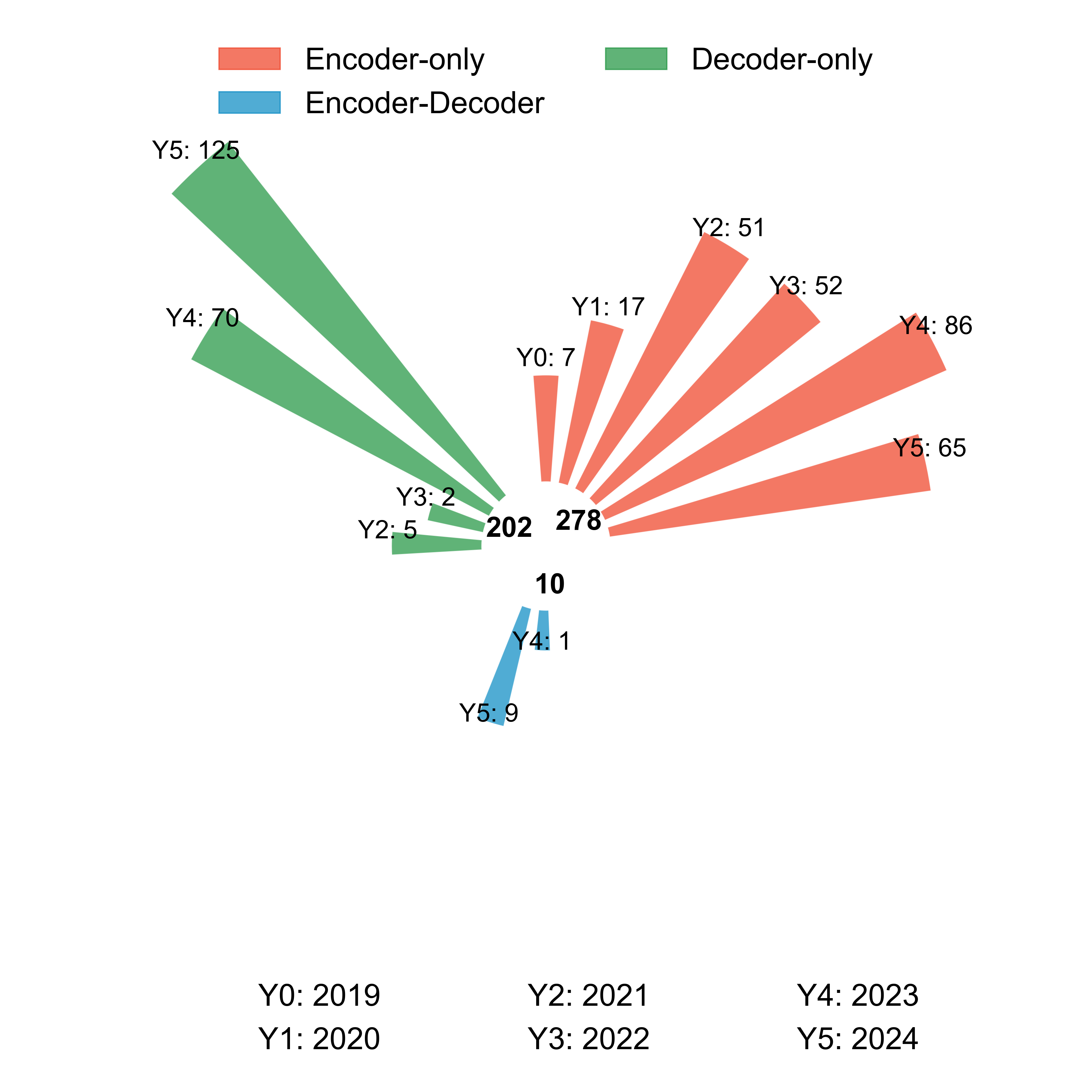}
\caption{Architectural trend. Temporal breakdown of encoder-only, decoder-only, and encoder-decoder LLM, reflecting the shifting architectural choices.}
\label{fig:data-processing22}
\end{minipage}
\begin{minipage}{1\linewidth}
\centering
\includegraphics[width=\linewidth]{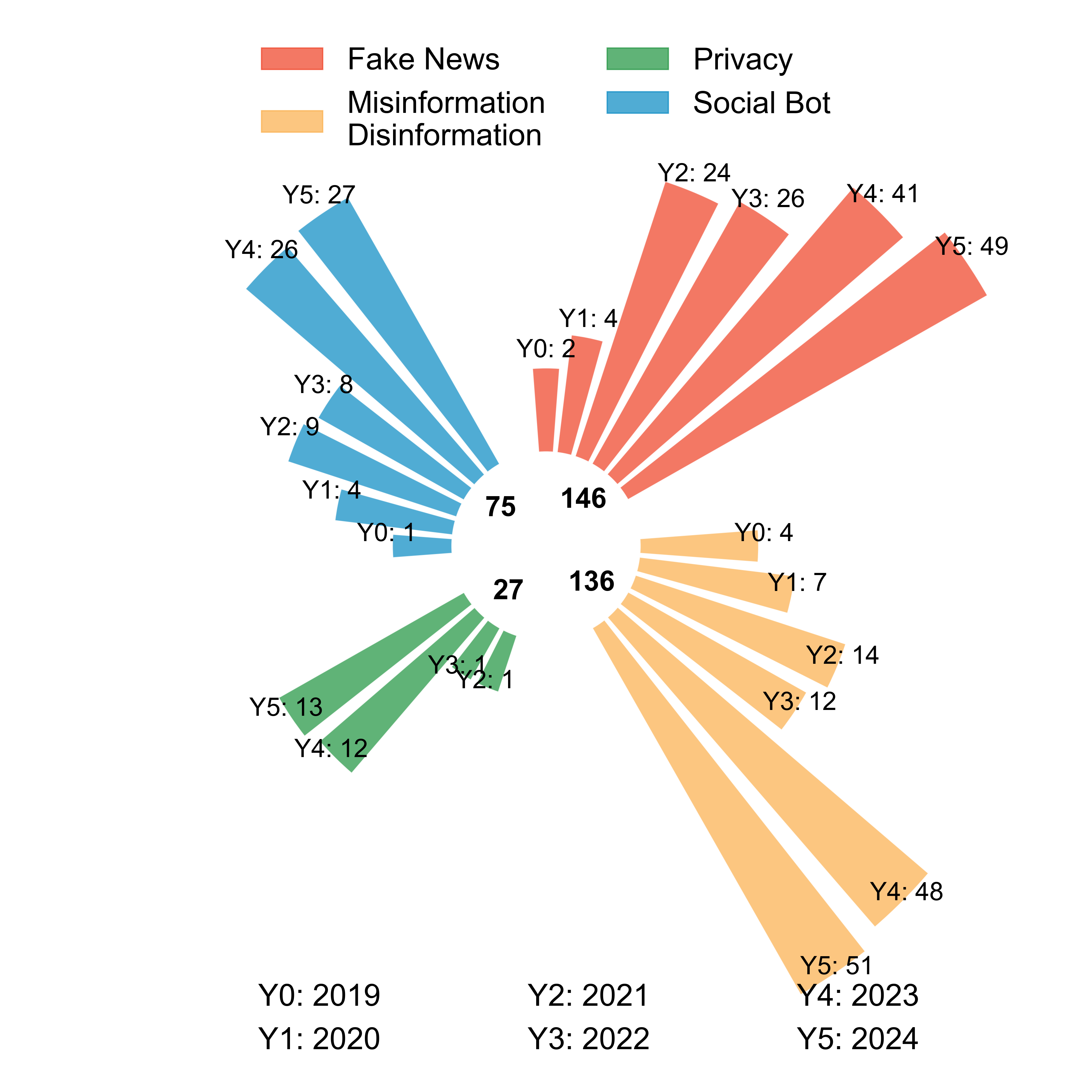}
\caption{Evolving research topics. Display of the annual distribution of trending topics in social media integrity research.}
\label{fig:data-processing33}
\end{minipage}
\vspace{-20pt}
\end{wrapfigure}

\textbf{Architectural Trends:} \textcolor{black}{The temporal distribution of model architectures shown in Figure \ref{fig:data-processing22} indicates an overall increase in reported LLM usage across encoder-only, decoder-only, and encoder–decoder paradigms within the sampled studies. Decoder-only architectures, such as the GPT series, appear frequently in recent work, likely reflecting their versatility in both content generation and analytical tasks. Encoder-only models (e.g., BERT and RoBERTa) continue to be represented, particularly in classification- and detection-oriented settings, where computational efficiency remains important. Encoder–decoder architectures (e.g., T5 and BART) are also present, offering hybrid capabilities that support tasks requiring both structured understanding and controlled generation. These architectural patterns should be interpreted as descriptive observations drawn from the analyzed corpus rather than as evidence of causal shifts in modeling practice.}


\textbf{Research Focus Distribution:} \textcolor{black}{The temporal analysis of research themes in Figure \ref{fig:data-processing33} highlights changes in the relative emphasis of different integrity-related topics over time within the sampled literature. While longstanding topics such as misinformation detection and content moderation continue to be widely studied, recent years show increased attention to emerging challenges, including advanced bot detection, multimodal deepfake analysis, and privacy-aware integrity mechanisms. These observed patterns reflect the evolving scope of integrity threats and corresponding research interests, though they are descriptive in nature and may be influenced by dataset availability, platform access, and broader shifts in the research ecosystem.}

\begin{figure}[t]
  \centering
  \includegraphics[width=0.6\textwidth]{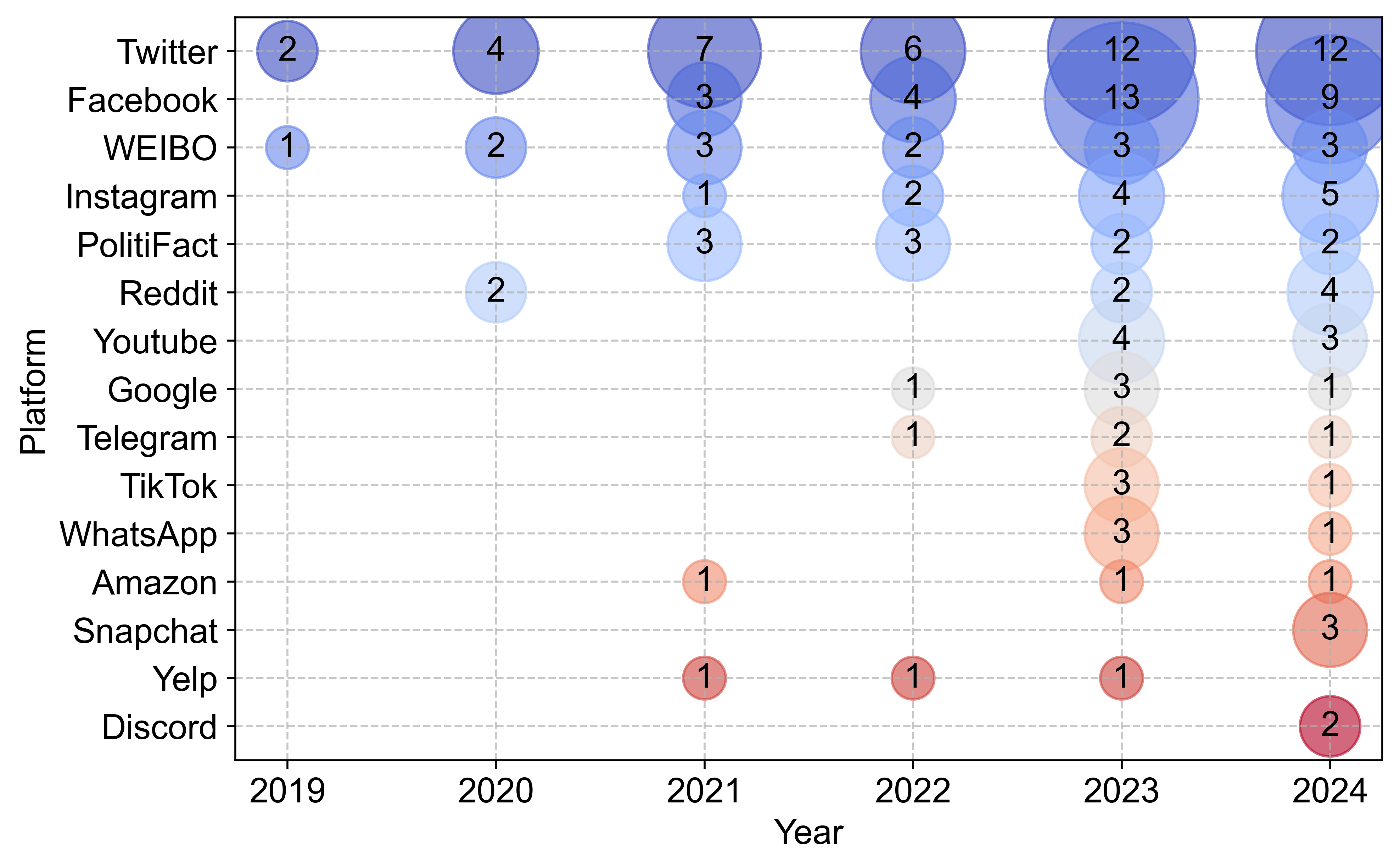}
  \captionof{figure}{\textbf{Temporal Analysis of LLM-based Social Media Integrity Research (2019-2024)} Platform-specific research distribution visualized through bubble charts, where bubble size indicates study frequency. The visualization reveals temporal trends and platform-specific focus in LLM-driven integrity research.}
  \label{fig:data-processing2}
  \vspace{-15pt}
\end{figure}

\textbf{Platform Distribution:} \textcolor{black}{The platform-level distribution illustrated in Figure \ref{fig:data-processing2} summarizes how LLM-based integrity research is represented across different social media platforms within the sampled studies. Twitter (now X) is the most frequently represented platform, which may be related to the availability of historical datasets and comparatively accessible APIs for studying misinformation and automated behavior. Other platforms, including Facebook and Reddit, maintain a steady presence, while platforms such as Discord and Mastodon appear less frequently but reflect emerging research interest in decentralized or community-driven social networks. These distributions should be interpreted in light of differences in data accessibility and platform policies rather than as comprehensive indicators of real-world platform impact.}

\color{black}
\subsection{Benchmark Datasets and Evaluation Metrics}

Evaluation practices in LLM-based security research are shaped by the heterogeneity of tasks, ranging from misinformation detection and bot identification to contextual privacy auditing. Thus, there is no single universal benchmark or evaluation metric. Instead, prior work has converged on a set of datasets and indicators that define the evaluation landscape. This subsection summarizes the benchmark datasets and evaluation metrics.

\textbf{Information Disorder:} Benchmark selection often matches the granularity of misinformation modeling. Claim-level veracity classification is commonly evaluated using LIAR~\cite{wang2017liar}, which provides short political statements with fine-grained truthfulness labels and standardized splits, making it a frequent testbed for both traditional and prompting-based LLM baselines~\cite{shu2020fakenewsnet, shu2017fake, gao2023rarr}. Article-level or discourse-level detection is often evaluated on FakeNewsNet~\cite{shu2020fakenewsnet}, which combines textual content with social engagement and propagation metadata, enabling evaluation of retrieval- or context-augmented models but introducing variability in pre-processing and feature usage. For domain-specific settings, health misinformation studies frequently adopt CoAID~\cite{cui2020coaid}, where robustness and domain transferability become central concerns.
As misinformation increasingly mixes text and images, multimodal benchmarks have become important~\cite{liummfakebench, papadopoulos2024verite}. Fakeddit is widely used due to its large-scale paired text-image data and multiple labeling granularities~\cite{nakamura2020fakeddit}, while newer datasets such as Mocheg emphasize cross-modal consistency rather than isolated classification~\cite{yao2023end}. Across these studies, metrics typically prioritize precision recall based indicators: accuracy is reported, but macro-F1 is often emphasized to address class imbalance~\cite{sundriyal2024crowd, shinde2025nlp, sharma2022construction}. When fact-checking includes rationale generation, works commonly add generation-oriented scores such as ROUGE~\cite{hu2025assessing}.

\textbf{Social Bot:} Bot detection benchmarking is relatively standardized. TwiBot-20~\cite{feng2021twibot} and TwiBot-22~\cite{feng2022twibot} are widely adopted as they provide large-scale labels together with behavior traces and graph-structured signals. Given strong class imbalance and deployment sensitivity, studies typically report precision, recall, and F1, often supplemented by AUC for ranking performance; accuracy alone is generally insufficient because small false-positive rates can translate into disproportionate moderation errors. Beyond correctness, some work also discusses reliability, using calibration metrics to reflect risks in threshold-based or human-in-the-loop pipelines.

\textbf{User Privacy:} Privacy and contextual integrity studies depart from standard label prediction. Rather than optimizing accuracy/F1, they assess whether an LLM discloses sensitive information in contexts where disclosure violates normative expectations~\cite{cai2023public, li2022dp, dickson2023ethics}. Benchmarks such as CONFAIDE define structured interaction scenarios and measure privacy failures as leakage or attack success rates under standardized prompting~\cite{tomekcce2024private}. Importantly, these evaluations often report utility--safety trade-offs: mitigations are assessed by both leakage reduction and the degree of performance degradation on benign prompts, reflecting that success is defined by minimizing harm rather than maximizing predictive accuracy alone.

\section{Structural Landscape and Emerging Challenges}
\begin{figure}[t]
     \centering
     \includegraphics[width=0.8\linewidth]{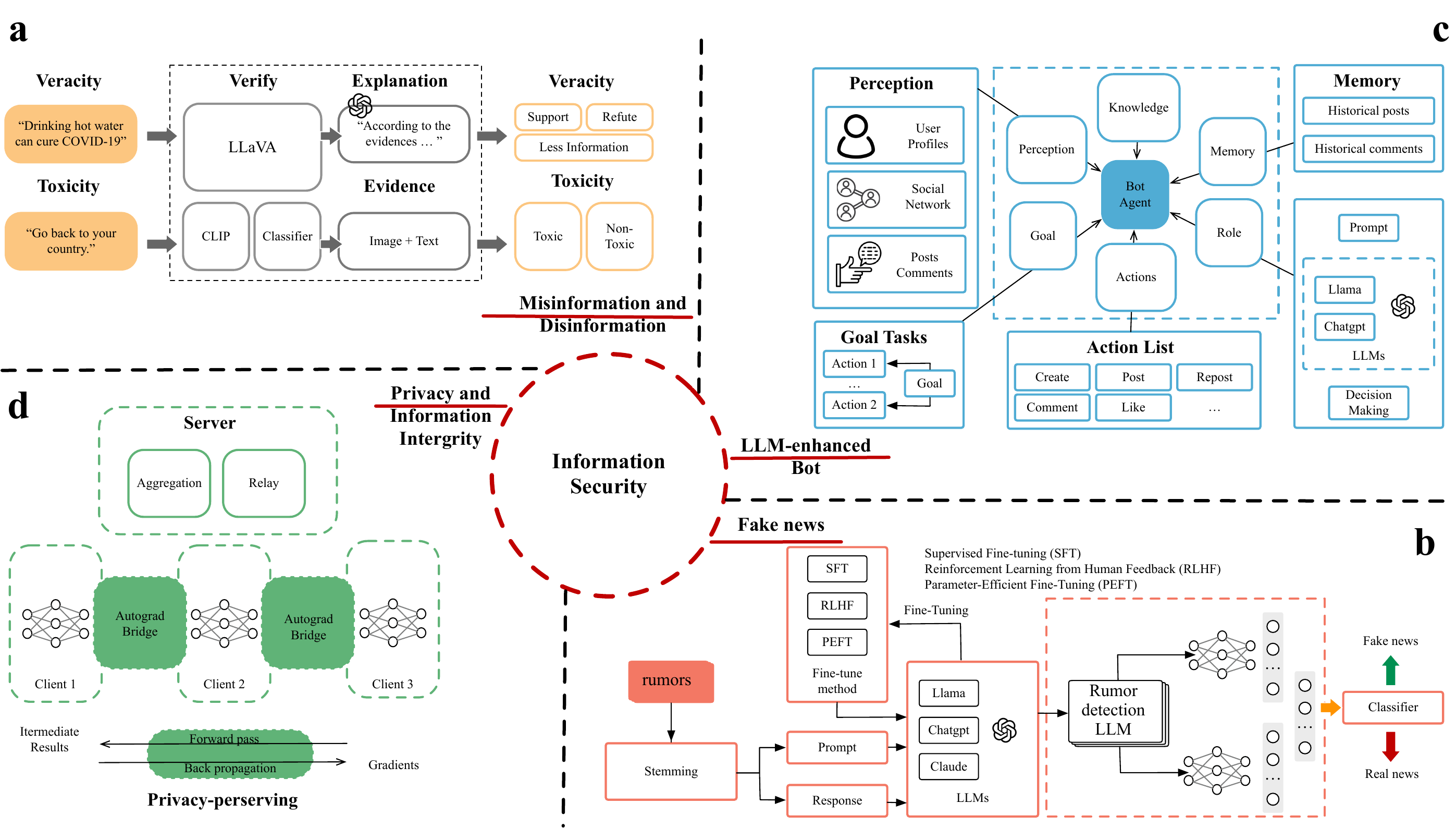}
     \caption{This figure illustrates four representative subdomain workflows (a–d) in information security research. (a)Misinformation and Disinformation: \citet{lee2024train} propose a multimodal knowledge transfer approach using LLaVA and CLIP to improve fact-checking and rumor detection. (b)Fake News: \citet{app14083532} develops automated detection systems for identifying and combating synthetic fake news content. (c)LLM-enhanced Bot Detection: \citet{Liu2023-gr} demonstrates advanced techniques for identifying automated accounts and coordinated bot networks. (d)Privacy and Information Integrity: \citet{su2024titanic} explores privacy-preserving mechanisms for protecting user data while maintaining information integrity in social media environments.}
     \label{fig: workflow}
     \vspace{-19pt}
 \end{figure}

As illustrated in Figure~\ref{fig: workflow}, the LLM-enabled social media integrity pipeline can be organized into four representative subdomain workflows, spanning misinformation detection, synthetic fake news analysis, bot governance, and privacy-preserving mechanisms. In the following, we examine these domains by distinguishing their emerging opportunities from their systemic challenges.

\subsection{Landscape: Opportunities in Social Media Information Integrity}

\begin{figure}[b]
    \centering
    \includegraphics[width=0.8\linewidth]{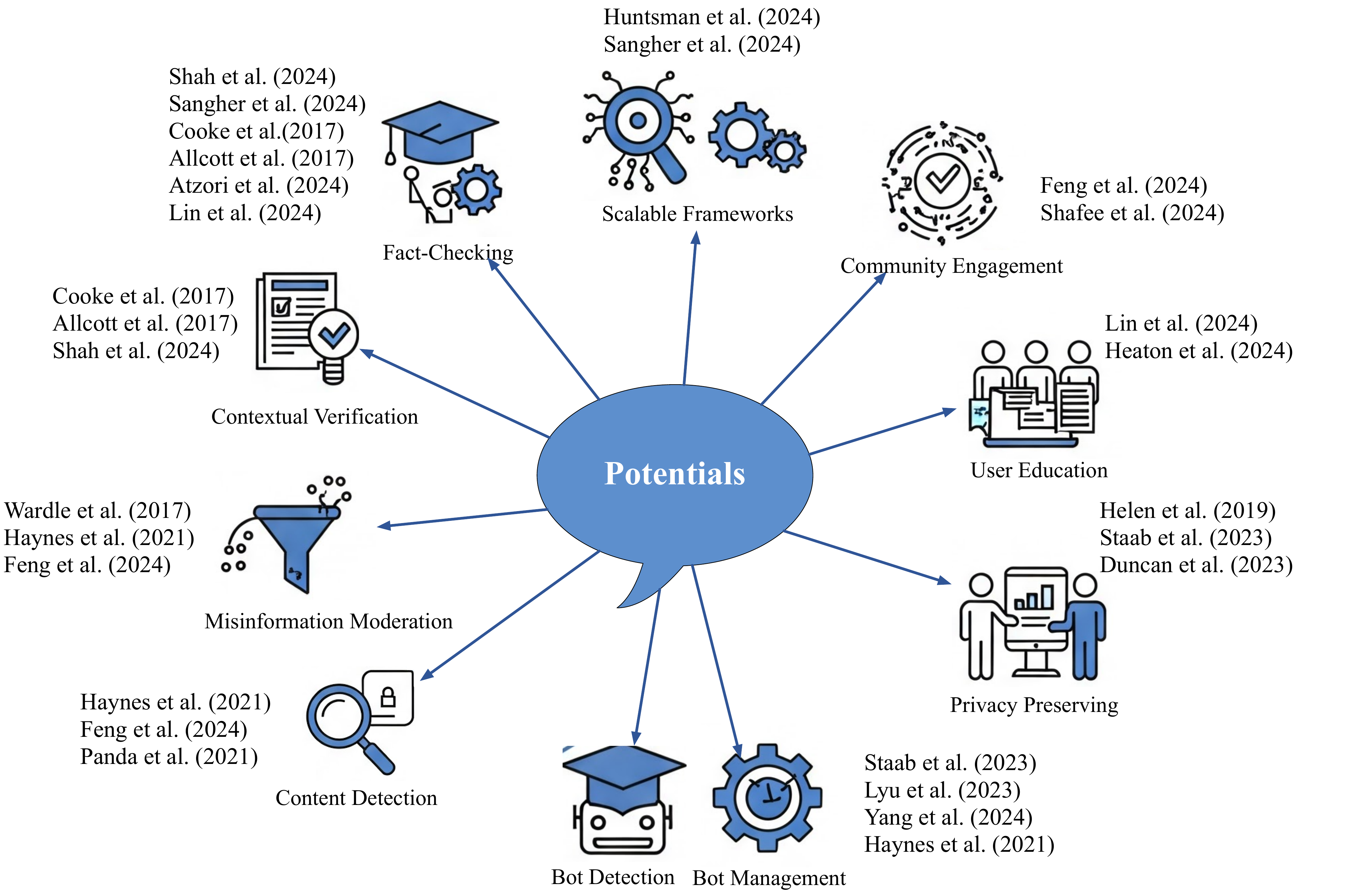}
    \caption{Potentials in LLMs for social media information integrity. 
    The figure illustrates key capabilities of LLMs in enhancing content verification, automated fact-checking, and content moderation, while also illustrating advances in privacy-preserving processing, user education, community engagement, and cost-effective deployment frameworks. Each directional component references recent research (e.g., Wardle et al. (2017) - \cite{wardle2017information}, Haynes et al. (2021) - \cite{Haynes2021-ga}, Panda et al. (2021) - \cite{Panda2021-bu}, Cooke et al. (2017) - \cite{cooke2017posttruth}, Allcott et al. (2017) - \cite{allcott2017social}, Shah et al. (2024) - \cite{Shah2024-js}, Atzori et al. (2024) - \cite{Atzori2024-tx}, Lyu et al. (2023) - \cite{lyu2023gpt}, Yang et al. (2024) - \cite{Yang2024-tq}, Duncan et al. (2023) - \cite{Duncan2023-xk}, Puppala et al. (2024) - \cite{Puppala2024-mx}, Nissenbaum et al. (2019) - \cite{nissenbaum2019contextual}, Lin et al. (2024) - \cite{Lin2024-xa}, Heaton et al. (2024) - \cite{Heaton2024-aw}, Staab et al. (2023) - \cite{staab2023beyond}, Feng et al. (2024) - \cite{Feng2024-gf}, Shafee et al. (2024) - \cite{Shafee2024-zc}, Panagiotou et al. (2021) - \cite{Panagiotou2021-uj}, Huntsman et al. (2024) - \cite{Huntsman2024-iw}, Sangher et al. (2024) - \cite{Sangher2024-pc})}
    \label{fig:Potentials}
\end{figure}

LLMs have delivered multi-dimensional gains for social-media information integrity, as illustrated in Figure~\ref{fig:Potentials}. When enhanced through domain-specific fine-tuning or retrieval-augmented generation (RAG), they achieve significant improvements in multilingual misinformation screening recall, reduce fact-checking latency through automated source verification, identify covert bot networks through pattern analysis, and generate privacy-aware prompts that strengthen user media-literacy. At the same time, careful system design, such as contextual-integrity guards, federated fine-tuning, and cost-aware routing, can keep these advances affordable and policy-compliant. In what follows, we summarize the landscape of LLM opportunities across key dimensions of social media information integrity.

{\textbf{Detect \& Moderate Misinformation.}} The proliferation of misinformation and disinformation on social media platforms represents one of the most pressing challenges in our digital age, with recent studies revealing that false information propagates as rapidly as truthful content~\cite{grinberg2019fake, tokita2024measuring}. Within this landscape, the advent of LLM marks a critical turning point, introducing a complex dynamic where these models serve both as potential solutions and possible sources of misinformation. While LLMs demonstrate remarkable capabilities in detecting inconsistencies across multiple languages and formats~\cite{dulhanty2019taking,Xu2024-rp,Zhang2024-ll}, their sophisticated generative abilities simultaneously present unprecedented challenges by producing highly convincing synthetic content~\cite{Borji2023-hb,spitale2023ai,Al-Omari2019-qg,Shah2024-js,yang2023rumor}. This dual nature of LLMs in the information integrity landscape requires careful examination of both their potential benefits and risks in social media environments, particularly as automated accounts continue to amplify the spread of both authentic and false information.
The challenge of protecting sensitive data while maintaining model utility presents a fundamental tension in LLM deployments. \citet{cai2023public} identify critical challenges in balancing analytical capabilities with privacy protection, particularly in mental health discussions. Traditional privacy-preserving techniques prove increasingly inadequate, as \citet{patsakis2023man} and \citet{mattern2022limits} show how LLMs can still infer user traits through subtle linguistic cues even after standard de-identification. This has led to exploration of more robust approaches, with \citet{li2022dp, huang2020instahide, wiseman2018learning} investigating differential privacy strategies, while \citet{su2024titanic} explores federated learning solutions for privacy-aware model updates. As \citet{martin2022ai} and \citet{dickson2023ethics} emphasize, there is an urgent need to develop stronger ethical guidelines and resolve the fundamental tension between analytical utility and privacy preservation.
{Emerging evidence suggests that LLM-enhanced detectors may improve recall by 6 to 10 percentage points over baseline BERT methods, though these results remain preliminary and require validation across diverse datasets and languages. \citet{lazer2018science, koka2024evaluating} report such gains on hate detector benchmark and misinformation detector benchmark, but the generalizability to real-world social media (such as class distributions, temporal drift, and adversarial attacks) remains an open question.} By modeling narrative intent, they distinguish inadvertent misinformation from deliberate disinformation, extending Wardle’s conceptual split \cite{wardle2017information}. This capability allows the same model to recognize both careless sharing and coordinated campaigns in diverse linguistic and cultural settings, even in low-resource languages \cite{Panda2021-bu}. \textcolor{black}{In the context of this survey, "coordinated networks" and "campaigns" are operationalized through specific detection methodologies found in the technical literature: For the community detection, to identify clusters of coordinated accounts, graph-based studies typically employ modularity optimization algorithms. The Louvain method is classically used to detect dense subgraphs, while recent works\cite{Chawla2023-le, feng2022twibot, shan2024multimodal, Yang2024-tq} increasingly adopt the Leiden algorithm to guarantee connected communities and improve clustering speed in large-scale social graphs. For the coordination heuristics: We distinguish between two primary detection signals: (1) Link-farm heuristics\cite{treves2023rurlman, Haynes2021-ga, Li2023-et}, which identify coordination through deterministic patterns such as shared URLs, identical posting times, or co-retweet structures; and (2) Probabilistic cascade models, which utilize causal diffusion analysis \cite{ijcai2024p873, shu2020fakenewsnet, jin2024veracity} to infer latent coordination by modeling the probability of information spreading between nodes, distinguishing organic viral spread from artificial amplification. 
Recent audits further demonstrate that LLMs outperform traditional filters in identifying nuanced health and political falsehoods, particularly those embedded in contextual ambiguity, or cross-domain narratives that rule-based and conventional classifiers systematically overlook \cite{Haynes2021-ga, Feng2024-gf}.
}

\textcolor{black}{Knowledge-augmented and retrieval-based detection pipelines differ substantially in how evidence is obtained and incorporated \cite{yang2023rumor,lee2024train,xuan2024lemma,zhou2024correcting}. In some settings, external evidence is provided in an oracle manner (e.g., curated knowledge or ground-truth supporting documents), while others rely on automated retrieval mechanisms such as BM25-style sparse search, dense vector retrieval, or API-based web search. These choices introduce trade-offs between evidence coverage and system latency, which are particularly important in real-time social media settings \cite{Chen2023-jf,zhou2024correcting}. Short-form content such as tweets or brief posts may also lead to failure modes where retrieval yields limited or ambiguous evidence.
In addition, retrieval-augmented systems vary in reasoning style \cite{xuan2024lemma,yang2023rumor,choi2024automated,cheung2023factllama}. Some adopt chain-of-thought generation to produce explanatory justifications, while others use verifier-style architectures that separately assess factual consistency. These design differences affect interpretability, computational cost, and deployment feasibility, highlighting that performance gains from knowledge-based detection should be understood within a broader system-level context rather than as purely model-level improvements.}
\textcolor{black}{Beyond architectural and retrieval design, the effectiveness of zero-shot and few-shot LLM-based detectors is also highly sensitive to inference configurations. Factors such as prompt template design, role framing, temperature, nucleus sampling ($p$), use of self-consistency, and tool-augmented reasoning chains can substantially influence model outputs. However, these settings are often inconsistently reported across studies, making direct performance comparisons difficult. 
We therefore argue that such inference parameters should be treated as part of the evaluation protocol rather than incidental implementation details. Explicit reporting of prompting and decoding configurations would allow future work to move toward more reproducible and head-to-head comparisons among LLM-based detection approaches.
}

As shown in Table \ref{tab:misinfo_techniques}, LLM-generated misinformation presents two distinct categories of challenges. The first is unintentional generation or hallucinations~\cite{Shah2024-js, Christensen2024-qq}, which often occur due to the intrinsic properties of LLMs. These hallucinations are particularly challenging, as they appear highly plausible in fine-grained details such as dates, names, and numbers, making them difficult to detect even when models attempt to generate factual content. Shah et al.~\cite{Shah2024-js} explore this dual role of LLMs in both the combating and the potential generating of misinformation, examining their impact on social media consumption and various domain applications. Christensen et al.~\cite{Christensen2024-qq} further demonstrate these challenges through specific cases like tourism planning.
The second category involves intentional generation - the deliberate misuse of LLMs for disinformation~\cite{Radivojevic2024-fj, barman2024dark}. This ranges from minor alterations to critical changes that dramatically alter meaning. Radivojevic et al.~\cite{Radivojevic2024-fj} demonstrate through experiments with GPT-4, LLama2, and Claude that humans can only identify the origin of such content 42\% of the time, with persona's influence exceeding human perception. Barman et al.~\cite{barman2024dark} further explore LLMs' potential in generating multimedia disinformation, emphasizing the need for ethical oversight and multi-stakeholder collaboration. These challenges are compounded by quality control issues, where verification must evaluate multiple aspects, including factual consistency and contextual alignment. 
The integration of these challenges with social media platforms creates additional complexities~\cite{dulhanty2019taking,jiang2020modeling,yang2023rumor}. The scalability and automation capabilities of LLMs enable rapid generation of false content that can be automatically adapted across platforms. This technological capability, combined with sophisticated audience-targeting and platform-specific optimization, significantly enhances the potential for echo chamber reinforcement. These issues are further complicated by biased datasets, adversarial attacks, and the risk of users taking automated fact-checking outputs as definitive truth without proper judgment~\cite{dulhanty2019taking,jiang2020modeling}. Additionally, as noted by Yang et al.~\cite{yang2023rumor}, data bias and topical bias in polarized comments remain underexplored challenges.

\begin{table*}[htbp]
\small
\centering
\caption{Representative Techniques of Fake News Detection.}
\label{tab:detection_techniques}

\resizebox{\textwidth}{!}{%
\begin{tabular}{l | p{2.5cm} | p{2.5cm} | p{2.2cm} | p{2.5cm} | p{2cm}}
\toprule
\textbf{Technique} 
& \textbf{Models} 
& \textbf{Scenario} 
& \textbf{Strength} 
& \textbf{Limitation} 
& \textbf{References} \\
\midrule

Transformer Learning
& BERT, RoBERTa, XLM-R
& Text-only;\newline cross-lingual
& Strong semantic modeling
& Domain shift;\newline weak grounding
& \cite{devlin2018bert,liu2019roberta,conneau2020unsupervised,chen2021transformer,malla2022fake,10.1093/llc/fqac049}
\\
\midrule

Explainability
& DistilBERT + SHAP
& Interpretable\newline support
& Improved\newline transparency
& Limited accuracy gain
& \cite{sanh2019distilbert,lundberg2017unified}
\\
\midrule

Multi-modal Fusion
& BDANN, FND-CLIP
& Text + image\newline detection
& Cross-modal\newline reasoning
& Sensitive to image quality
& \cite{9206973,zhou2022multimodalfakenewsdetection,cmc.2024.046202}
\\
\midrule

Prompt-based Learning
& ChatGPT,\newline GenFEND
& Zero-/few-shot
& No task-specific training
& Prompt \& decoding sensitivity
& \cite{fi16080286,nan2024let,huang2023fake}
\\
\midrule

Few-shot Learning
& RumorLLM
& Low-resource settings
& Reduced labeling cost
& Unstable with tiny data
& \cite{app14083532}
\\
\midrule

Privacy-preserving
& AugFake-BERT
& Cross-organization
& Privacy\newline compliance
& Performance\newline trade-off
& \cite{app12178398}
\\
\midrule

Domain Transfer
& CT-BERT, RoBERTa
& Region/domain adaptation
& Better\newline generalization
& Needs domain\newline pretraining
& \cite{malla2022fake,stipiuc2024romanian}
\\

\bottomrule
\end{tabular}%
}

\vspace{-10pt}
\end{table*}

\begin{table}[htbp]
\small
\centering
\caption{Representative Techniques of Misinformation.}
\label{tab:misinfo_techniques}

\resizebox{\linewidth}{!}{%
\begin{tabular}{l | p{1.8cm} | p{2.8cm} | p{1.8cm} | p{3cm} | p{1.5cm}}
\toprule

\textbf{Model / Method} 
& \textbf{Supervision} 
& \textbf{Knowledge Integration} 
& \textbf{Modality} 
& \textbf{Setting} 
& \textbf{References} \\

\midrule

ChatGPT / LLaMA 
& Zero-shot 
& Retrieval-augmented 
& Text + Vision 
& Online / Streaming 
& \cite{Chen2023-jf} \\

\midrule

FACT-GPT / FactLLaMA 
& Fine-tuned 
& Knowledge graphs + Retrieval 
& Text 
& Online / Streaming 
& \cite{cheung2023factllama,choi2024automated,lee2024train} \\

\midrule

LEMMA 
& Few-shot 
& Retrieval / Tool-use 
& Text 
& Offline / Batch 
& \cite{xuan2024lemma,yang2023rumor} \\

\midrule

LLaVA 
& Few-shot 
& Retrieval or Transfer 
& Multimodal 
& Offline / Batch 
& \cite{Radivojevic2024-fj} \\

\bottomrule
\end{tabular}%
}

\vspace{-8pt}
\end{table}


\begin{table}[htbp]
\small
\centering
\caption{LLM Applications and Their Privacy Implications: A Comprehensive Overview of Privacy-preserving Techniques in Social Media Contexts}
\label{tab:llm_privacy}

\resizebox{\linewidth}{!}{%
\begin{tabular}{l | p{4.5cm} | p{4.6cm} | p{1.5cm}}
\toprule

\textbf{Techniques} 
& \textbf{Methods} 
& \textbf{Description} 
& \textbf{References} \\

\midrule

Knowledge Distillation
& Privacy-aware model compression and transfer
& Minimize sensitive data exposure during training
& \cite{zhao2021knowledge,filippo2024future} \\

\midrule

Differential Privacy
& DP-SGD and privacy-preserving optimization
& Formal privacy guarantees for model training
& \cite{li2022dp,huang2020instahide} \\

\midrule

Federated Learning
& Decentralized model updates and training
& Privacy-preserving distributed learning
& \cite{su2024titanic,yao2019federated} \\

\midrule

Adversarial Protection
& Text perturbation and obfuscation techniques
& Prevent unauthorized inference and identity linking
& \cite{dogan2023catch,supriya2022p} \\

\midrule

Privacy-Preserving Attention
& Modified attention mechanisms for privacy
& Secure information processing in LLMs
& \cite{carlini2021extracting,patsakis2023man} \\

\bottomrule
\end{tabular}%
}

\vspace{-8pt}
\end{table}

\textcolor{black}{{\textbf{Political Economy and Platform Incentives}}
The dynamics of misinformation spread cannot be fully understood from a purely technical perspective. These dynamics are influenced by platform business models, governance choices, and resource asymmetries among various actors. User engagement ranking algorithms prioritize content that generates clicks, shares, or increases user dwell time, creating selective pressure that favors sensational, polarizing, or emotionally charged content, which LLMs can efficiently generate or amplify \cite{allcott2017social, tufekci2017twitter}. Furthermore, the high fixed costs of training and deploying state-of-the-art LLMs concentrate defensive capabilities in the hands of well-resourced platforms and vendors, creating an asymmetry between attackers and defenders \cite{benkler2018network}. These structural dynamics mean that purely technical mitigation measures will be insufficient unless accompanied by changes in incentive structures. Therefore, current researchers call for interdisciplinary research integrating fields such as system design, economics, and governance to assess how different incentive structures alter the generation, detection, and dissemination of misinformation amplified by LLMs.}
\textcolor{black}{{{Second-order Effects of LLM-based Moderation}} 
While LLMs can improve moderation coverage and responsiveness, deploying them at scale may introduce unintended societal harms. Over enforcement or low threshold automation can create a chilling effect by increasing false positives on legitimate but controversial speech, thereby narrowing the range of acceptable discourse. In addition, moderation systems trained primarily on dominant linguistic and cultural data may systematically mis-handle minority dialects, producing disparate error rates and cultural homogenization. These risks motivate safeguards such as calibrated thresholds and subgroup robustness auditing alongside conventional accuracy metrics.}

{\textbf{Multimodal Fact-Checking \& Contextual Verification.}}
\textcolor{black}{Preliminary evidence indicates that coupling GPT-4 claim decomposition with retrieval augmented generation could reduce fact-checker latency (reported as ~35\% improvement in \citet{russo-etal-2023-benchmarking}). However, this result is based on controlled experiments and may not generalize to real-world social media fact-checking pipelines, which face challenges such as API latency variability, content diversity, and heterogeneous crowd worker pools. Agentic decomposition approaches (\emph{e.g.}, FactAgent \cite{Feng2024-gf}) similarly show potential but require further evaluation regarding scalability, worker quality consistency, and cost effectiveness.} Cross-platform context analysis uncovers temporal or source inconsistencies that headline-only systems overlook \cite{Shah2024-js}, and when these pipelines incorporate news format, presentation style, and narrative evolution tracking, they can flag emerging misinformation trends earlier in the news cycle \cite{Sangher2024-pc}. Recent studies report macro-F1 scores around 0.82 on real-time news streams \cite{10.1093/llc/fqac049}, surpassing traditional fact-checking workflows \cite{cooke2017posttruth, allcott2017social, Atzori2024-tx, Lin2024-xa}. 

Table~\ref{tab:detection_techniques} presents a systematic overview of fake news detection techniques, illustrating their evolution from basic transformer models to sophisticated multi-modal and domain-specific applications. Among these techniques, transformer-based learning forms the foundation of modern fake news detection, with models like BERT and RoBERTa demonstrating strong performance in text analysis. First, basic models like BERT, RoBERTa, and XLM-RoBERTa excel in text-based detection, with fine-tuned RoBERTa models specifically targeting COVID-19 misinformation~\cite{chen2021transformer,malla2022fake}. Second, cross-lingual detection capabilities are enhanced through XLM-RoBERTa-CNN combinations~\cite{10.1093/llc/fqac049}. Third, hybrid models like BDANN integrate BERT with VGG-19 for initial multi-modal detection~\cite{9206973}, while DistilBERT improves explainability~\cite{ayoub2021combat}. 
Advanced multi-modal detection has evolved through three key innovations. First, FND-CLIP and EANBS process text and visual inputs simultaneously through contrastive learning and integrated BERT-CNN architectures~\cite{zhou2022multimodalfakenewsdetection,cmc.2024.046202}, enabling the detection of inconsistencies between textual claims and associated images. Second, GPT-4 enhances detection by identifying subtle cross-modal inconsistencies~\cite{lyu2023gpt}, particularly effective in cases where text and images have been manipulated to create misleading narratives. The model's strong performance in cross-modal reasoning helps identify cases where images have been repurposed with false textual contexts or where subtle visual manipulations contradict textual claims.
Additionally, generative frameworks have advanced detection capabilities in two ways: ChatGPT enables zero-shot classification in low-resource settings~\cite{fi16080286}, allowing effective detection even without extensive training data, while GenFEND enriches training data through scenario simulation~\cite{nan2024let}, creating diverse misinformation patterns.
Domain Transfer and Cross-lingual Learning approaches address the challenge of adapting detection models across different domains and languages. CT-BERT demonstrates effective domain adaptation for COVID-19 misinformation~\cite{malla2022fake}, while RoBERTa-based models show promising results in cross-lingual transfer for Romanian political misinformation detection~\cite{stipiuc2024romanian}. These domain transfer techniques are particularly crucial when dealing with emerging topics or low-resource languages. For instance, ChatGPT-based transfer learning has shown effectiveness in adapting bias detection across different cultural contexts~\cite{huang2023fake}. Meanwhile, complementary techniques like active learning in RumorLLM~\cite{app14083532} and federated learning in AugFake-BERT~\cite{app12178398} enhance the efficiency and privacy of these domain adaptation processes.
Beyond detection, LLMs advance fake news understanding through simulation and analysis. The Fake News Propagation Simulation framework studies diffusion patterns and virality factors~\cite{ijcai2024p873}, while Fact Agent models analyze dissemination patterns and develop real-time mitigation strategies~\cite{li2024large}.

{\textbf{Bot Detection, Analysis \& Management.}}  
LLM-enhanced detectors jointly model textual semantics, image cues, temporal ``digital-DNA'' traces, and interaction graphs, yielding up to 9 percentage point gains in F1 score on the TwiBot-22 benchmark \cite{feng2022twibot, staab2023beyond, lyu2023gpt}. By analyzing behavioral patterns, content-generation fingerprints, and network dynamics, they surface coordinated, cross-platform inauthentic activity \cite{Yang2024-tq,allcott2017social}. These systems can also separate sophisticated social-engineering bots from legitimate automation \cite{Haynes2021-ga}, enabling fine-grained intervention policies that uphold overall platform integrity. 
Research into using LLM-enhanced bots for maintaining information integrity has evolved significantly, with substantial advances in detection and prevention capabilities.
Early detection systems established fundamental approaches that continue to influence current solutions. Dukić et al. \cite{Dukic2020-ig} and Heidari et al. \cite{Heidari2020-xk} pioneered the integration of emoji embeddings and sentiment features with BERT models, demonstrating significant improvements in bot detection accuracy. These foundational works were extended by Xu et al. \cite{Xu2021-pl} through the combination of ALBERT with Bi-LSTM and self-attention mechanisms, establishing new benchmarks in bot spam detection performance.
Multilingual detection capabilities have seen remarkable advancement. A series of studies demonstrated effective cross-lingual approaches. Panda et al. \cite{Panda2021-bu} developed models for detecting misinformation across English, Bulgarian, and Arabic; Garcia-Diaz et al. \cite{Garcia-Diaz2022-nu} achieved over 97\% accuracy in Spanish language satire detection; and Milkova et al. \cite{Milkova2023-gw} successfully detected value-expressive posts in Russian social media. These works collectively demonstrate the adaptability of LLM-based bot detection systems across different linguistic contexts.
Advanced architectural approaches have significantly improved detection capabilities. Kumar et al. \cite{Kumar2021-rf} and Guo et al. \cite{Guo2021-jf} developed hybrid frameworks combining BERT with graph convolutional networks, while Garcia-Silva et al. \cite{Garcia-Silva2021-hb} demonstrated the superiority of generative transformers in bot detection tasks. Recent work by Ghanem et al. \cite{Ghanem2023-ku} and Raga et al. \cite{Raga2022-sh} has pushed performance boundaries further, achieving accuracy rates above 98\% across different platforms. \textcolor{black}{Although several studies report detection accuracies exceeding 98\% across different platforms, such results should be interpreted with caution. Bot detection benchmarks are particularly sensitive to evaluation leakage, where structural or temporal overlap between training and test data may inadvertently inflate performance. For instance, shared retweet networks, synchronized activity bursts, or partially overlapping account clusters can allow models to learn dataset-specific signatures rather than generalizable bot behaviors.}
Real-time monitoring and prevention bot systems have emerged as crucial defensive tools. Panagiotou et al. \cite{Panagiotou2021-uj} developed News Monitor, achieving high accuracy in rumor identification, while Kawintiranon et al. \cite{Kawintiranon2022-ex} and Catelli et al. \cite{Catelli2022-fv} demonstrated effective approaches for context-specific bot spam detection. These systems have been enhanced by recent work from Ramamonjisoa et al. \cite{Ramamonjisoa2024-oh} and Bonechi et al. \cite{Bonechi2024-zi}, who developed sophisticated automated moderation systems for bots.
Specialized detection approaches have shown promising results in specific domains. Jones et al. \cite{Jones2023-sf} and Sallah et al. \cite{Sallah2024-qg} developed bot systems specifically targeting child protection on social media, while Riyantoko et al. \cite{Riyantoko2022-dp} demonstrated effective SMS bot spam classification using combined LSTM and BERT. 

{\textbf{Privacy-Preserving Content Processing.}}
The integration of LLMs into social media platforms has fundamentally reshaped the privacy landscape. With their ability to infer and memorize personal details across modalities, LLMs introduce privacy risks far beyond traditional NLP systems. Table~\ref{tab:llm_privacy} summary current privacy-preserving techniques and their applications in social media contexts. 

\textbf{Privacy threat model:}
\textcolor{black}{In social media platforms deploying LLMs, privacy risks arise from specific threat actors and attack surfaces. Key assets include user personal data, private messages, behavioral profiles, and sensitive group attributes. Adversaries may include external attackers probing model APIs, malicious insiders with training data access, or coordinated actors exploiting prompt-based leakage. Typical attack surfaces include training data collection pipelines, model outputs, retrieval components, and user facing conversational interfaces. Core threats include membership inference, model inversion, contextual integrity violations, and cross session leakage, with risk amplified under real time interaction and large scale personalization.}

\textcolor{black}{
To make privacy risks actionable in LLM-enabled social media systems, we map core threats to concrete controls. Membership inference and training data exposure motivate privacy-preserving training such as DP-SGD and data minimization. \textcolor{black}{ Differential Privacy solutions, such as DP-SGD, are considered the standard for providing $(\epsilon, \delta)$-DP guarantees. Its efficacy in social media contexts is governed by the clipping normalization, typically calibrated between $[0.1, 1.0]$, to bound the influence of outlier "heavy-hitter" users while minimizing the convergence floor of training loss curves. Similarly, Federated Learning (FL) solutions address decentralized storage but must incorporate Byzantine-robust aggregation. These developments address the need for context-aware privacy reasoning while maintaining a Pareto-optimal utility-privacy frontier. } Model inversion and sensitive attribute inference require output filtering, regularization, and red-team leakage testing. Contextual leakage in conversational settings motivates contextual integrity constraints and prompt-level guardrails, though adaptive jailbreaks remain a residual risk. Cross-session leakage and personalization risks motivate isolation mechanisms for memory and retrieval, with continuous monitoring of leakage rates over long context interactions. These mappings point out that privacy governance must combine technical mitigation with ongoing auditing and deployment time enforcement.}
\textcolor{black}{Grounded in Nissenbaum’s contextual-integrity theory \cite{nissenbaum2019contextual}, modern LLM pipelines aim to deliver personalized curation while respecting user privacy. Audits such as LLM-CI show that vanilla prompting can leak context-specific private attributes in a substantial fraction of cases (e.g., around 39\%), underscoring the difficulty of enforcing contextual privacy guarantees in practice. Some studies explore embedding contextual-integrity–aware constraints as a mitigation strategy and report reduced leakage under specific experimental settings, though the generality of these gains and their utility trade-offs remain to be systematically validated \cite{lan2025contextual}.} Complementary safeguards include differential-privacy masking and other privacy-preserving analytics \cite{staab2023beyond}, transparent data-handling and audit trails \cite{Duncan2023-xk}, and consent-aware model updates via federated or split learning \cite{Puppala2024-mx}. Together these measures enable platforms to offer fine-grained, trustworthy recommendations while upholding robust privacy guarantees.

{\textbf{User Education \& Media-Literacy Support.}}
\textcolor{black}{Instruction-tuned LLMs have been explored as tools for generating adaptive ``trust cues,'' critical-thinking scaffolds, and media-literacy interventions. In controlled user studies, one line of work reports improvements in misinformation-recognition accuracy on the order of 12\% under specific experimental settings \cite{Heaton2024-aw}, though the generalizability and durability of these effects remain open questions.} Beyond static tips, the models dynamically tailor explanations and verification checklists to a user’s demonstrated knowledge and engagement level, drawing on multilingual value-alignment work that transfers across languages with little performance loss \cite{huang-etal-2024-flames}.  They also incorporate practical guidance for source checking and claim verification \cite{Lin2024-xa}, thereby empowering users to recognize manipulation techniques and exercise informed skepticism.
{\textbf{Community Engagement \& Collaborative Verification.}}
LLMs catalyze constructive dialogue by generating prompts and replies that nudge users toward evidence-based discussion \cite{staab2023beyond}.  In civic fact-checking forums, such prompts have doubled the volume of user-submitted evidence while avoiding alert-fatigue patterns \cite{10.1093/llc/fqac049}. Agentic pipelines like FactAgent decompose verification into micro-tasks that crowd workers validate asynchronously \cite{Feng2024-gf}, yet still manage resource constraints \cite{Shafee2024-zc} and platform-integration complexity \cite{Ramamonjisoa2024-oh}.

\subsection{Landscape: Challenges in Social Media Information Integrity}

\begin{figure}[b]
    \centering
    \includegraphics[width=0.8\linewidth]{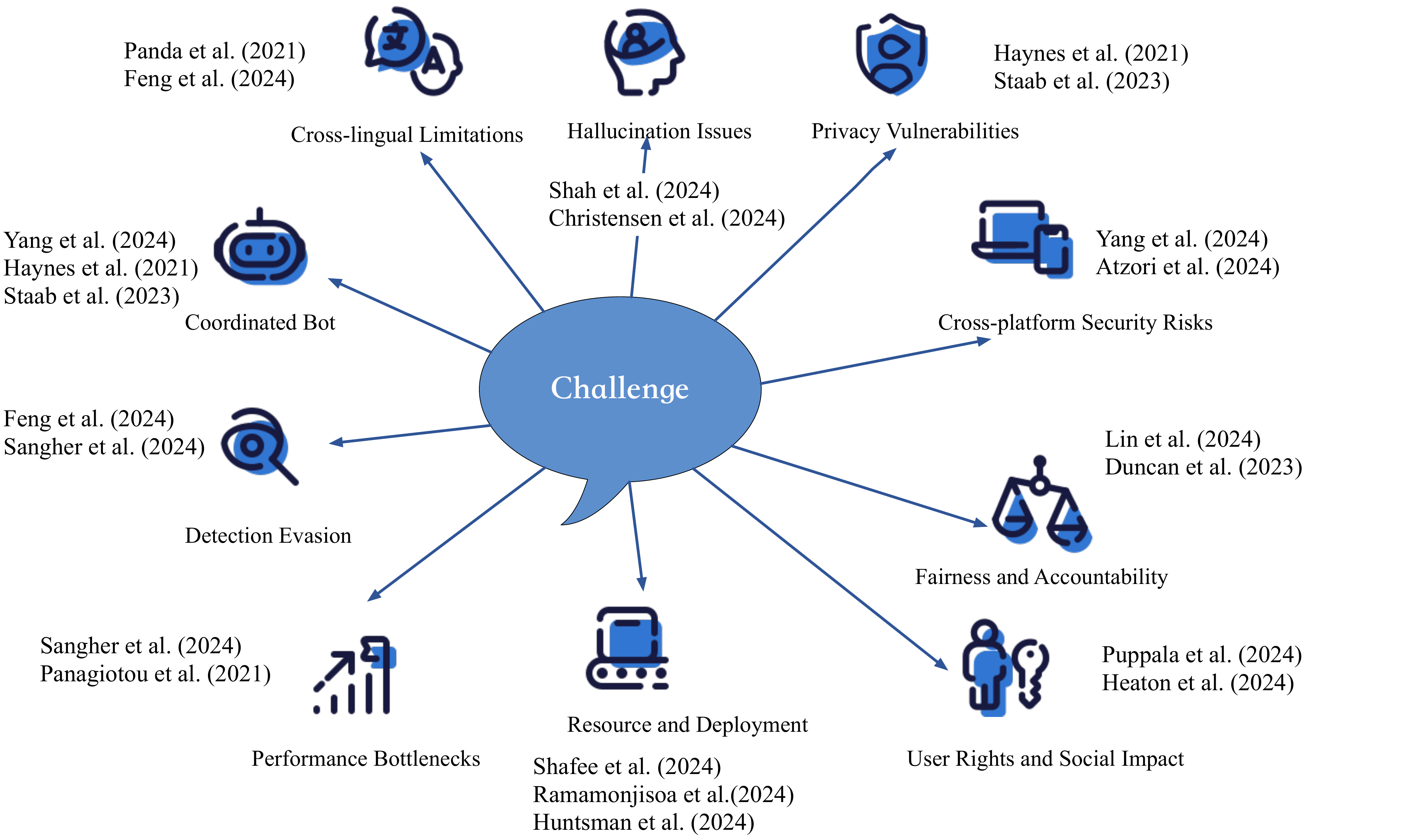}
    \caption{Challenges in LLMs for social media information integrity. 
    The figure illustrates the key capabilities of LLMs in privacy vulnerabilities, cross-platform security risks, and fairness, as well as accountability, user rights, social impact, resource and deployment, performance bottlenecks, detection evasion, coordinated bots, and cross-lingual limitations. Each directional component references recent research (e.g., Feng et al. (2024) – \cite{Feng2024-gf}, Sangher et al. (2024) – \cite{Sangher2024-pc}, Panda et al. (2021) – \cite{Panda2021-bu}, Shah et al. (2024) – \cite{Shah2024-js}, Haynes et al. (2021) – \cite{Haynes2021-ga}, Staab et al. (2023) – \cite{staab2023beyond}, Yang et al. (2024) – \cite{Yang2024-tq}, Atzori et al. (2024) – \cite{Atzori2024-tx}, Lin et al. (2024) – \cite{Lin2024-xa}, Duncan et al. (2023) – \cite{Duncan2023-xk}, Heaton et al. (2024) – \cite{Heaton2024-aw}, Puppala et al. (2024) – \cite{Puppala2024-mx}, Shafee et al. (2024) – \cite{Shafee2024-zc}, Ramamonjisoa et al. (2024) – \cite{Ramamonjisoa2024-oh}, Panagiotou et al. (2021) – \cite{Panagiotou2021-uj}, Huntsman et al. (2024) – \cite{Huntsman2024-iw}.
)}
    \label{fig:Challenges}
\end{figure}

Our landscape analysis also reveals four primary challenge domains in LLM-based social media information integrity, as illustrated in Figure~\ref{fig:Challenges}. These findings emphasize critical concerns across technical implementation, ethical considerations, deployment feasibility, and user interaction.

{\textbf{Detection and Verification Challenges}} Technical obstacles in LLM-based information integrity now center on model-specific detection evasion, inference performance bottlenecks, and hallucination issues. Detection evasion remains a foremost obstacle: adversaries continually refine prompt engineering tactics to bypass LLM integrity filters, with evasion success rates reported as high as 29.6\% \cite{Feng2024-gf}. Performance bottlenecks in LLM inference and prompt processing continue to hinder timely and accurate integrity verification across high-volume social media streams \cite{Feng2024-gf, Sangher2024-pc}. LLM cross-lingual capabilities remain limited, particularly in detecting misinformation across non-English or low-resource languages \cite{Panda2021-bu}. Most critically, model hallucination undermines trust in LLM-generated content, as spurious outputs can evade verification protocols and propagate misinformation \cite{Shah2024-js}.


\color{black}
{\textbf{Resource and Deployment Challenges}} LLM deployment for security and integrity applications faces significant challenges in resource management, system integration, and long-term sustainability. The substantial computational demands of LLMs, combined with complex integration requirements, impede seamless deployment and maintenance at scale, particularly for systems that rely on continuous monitoring or multi-stage reasoning pipelines \cite{Shafee2024-zc, Ramamonjisoa2024-oh, Huntsman2024-iw}. Beyond technical integration, recurring inference costs and energy consumption introduce additional economic burdens, which can limit the feasibility of sustained deployment for organizations with constrained resources \cite{strubell2019energy, patterson2021carbon}.
These challenges are further compounded by LLM-specific operational factors, including prompt optimization, model quantization, and the need to maintain multiple model variants across heterogeneous integrity tasks \cite{Panagiotou2021-uj, Ramamonjisoa2024-oh}. Importantly, disparities in access to computational infrastructure and proprietary models can exacerbate global inequalities, making LLM-based defenses less accessible to institutions in low-income or Global South regions despite comparable security needs~\cite{StanfordHAI2024-ai}.

{\textbf{Privacy and Security Challenges}} LLM-specific security risks center on prompt injection vulnerabilities, training data privacy, and model extraction threats. The privacy issues are particularly concerning for LLMs, as these models may memorize and inadvertently leak sensitive information from their training data, while also being vulnerable to advanced prompt engineering exploits~\cite{xiong-etal-2025-invisible, Haynes2021-ga, staab2023beyond}. Furthermore, these challenges are exacerbated by issues related to data provenance, user consent, and content ownership, particularly when LLMs are trained on or applied to large-scale social media data collected through scraping. Even when the content on this social media may be publicly accessible, its reuse for model training or downstream security analysis raises concerns about permissible use, anonymization, and compliance with data protection regulations such as GDPR~\cite{GDPR} and CCPA~\cite{CCPA}. \textcolor{black}{While privacy-preserving techniques such as Differential Privacy (DP) and Federated Learning (FL) provide important safeguards against attacks like model inversion and membership inference, they introduce non-negligible trade-offs in large-scale social media settings. DP mechanisms typically add noise to gradients or outputs, which can degrade model utility and require additional training iterations to maintain performance. Similarly, FL shifts computation to distributed clients and introduces communication overhead, synchronization delays, and potential model divergence across heterogeneous devices.
These constraints are particularly relevant in the context of real-time social media monitoring, where high-throughput and low-latency processing are critical. LLM-based systems already face substantial computational bottlenecks; integrating DP or FL can further increase latency and resource consumption. Therefore, while these privacy-preserving approaches are promising, their deployment in real-time misinformation or bot detection systems requires careful balancing between privacy guarantees and operational performance.}
In addition, adversaries can leverage LLMs to generate highly convincing phishing content and automate social engineering attacks at scale~\cite{Yang2024-tq}. The integration of LLMs across multiple platforms further complicates security and compliance, as attackers can exploit variations in model behavior, data handling practices, and regulatory enforcement across services~\cite{Atzori2024-tx}. Therefore, these challenges demonstrate that privacy-aware training pipelines, transparent data governance, and consent-sensitive deployment strategies are necessary when integrating LLMs into security-critical applications.

{\textbf{Fairness and Accountability Challenges}} 
Ethical considerations specific to LLMs  also encompass fairness, transparency, explainability, and accountability in safety-critical deployments. Challenges related to fairness stem from biases embedded in training data, model architectures, and evaluation benchmarks. These biases may cause LLMs to perpetuate or amplify societal prejudices during content generation and moderation, thereby disproportionately affecting marginalized groups [Lin2024-xa, Duncan2023-xk]. In social media contexts, such biases can lead to inconsistent moderation outcomes across different languages, dialects, and cultural backgrounds, directly influencing user participation, visibility, and inclusiveness.
The opacity and probabilistic nature of LLMs further complicate accountability. Many LLM-based moderation and content integrity systems operate as black boxes, making it difficult to trace decision-making processes or to explain why specific content is flagged, suppressed, or promoted. This lack of transparency can undermine user trust, hinder effective appeals, and complicate external audits by platform operators or regulatory bodies. Although post hoc explanation techniques—such as rationale generation or attention-based visualizations—show some promise in improving explainability, they may not faithfully reflect the underlying model behavior, raising concerns about the reliability of explanations and the risk of “explanation washing.”
These challenges are magnified when LLMs are integrated into complex, multi-stage pipelines that involve multiple models, third-party APIs, and platform-specific policies. In such settings, accountability is diffuse: responsibility for erroneous or biased outcomes may be unclear, and tracing failures across components becomes difficult. Addressing these issues requires explicit accountability mechanisms, including decision traceability, transparent logging, and well-defined human-in-the-loop oversight~\cite{Heaton2024-aw, Puppala2024-mx}.

\color{black}



\section{Discussion}

 


\subsection{Key Findings}

\textbf{Leading Models and Research Paradigms.}
Based on our review, the field of LLM-driven social media information integrity has crystallized around several influential models and research approaches.
First, in fact-checking and misinformation detection, LLM-based models like FACT-GPT~\cite{choi2024automated} and FactLLaMA~\cite{cheung2023factllama} have emerged as industry standards. These models excel at automated verification but face a critical challenge: they are prone to hallucinations in zero-shot and cross-domain scenarios, particularly when verifying specific details like dates, numbers, or named entities~\cite{Shah2024-js, Christensen2024-qq}.
In the privacy and security domain, frameworks like ProPILE~\cite{kim2024propile} and P-BERT~\cite{supriya2022p}, as well as federated learning approaches such as SocFedGPT~\cite{puppala2024socfedgpt}, have emerged as leading solutions. While these LLM-based models have advanced detection capabilities, they also expose critical challenges in generalizability, interpretability, and robustness.

\textbf{Trends and Future Breakthroughs.}
\textcolor{black}{The research frontier is increasingly oriented toward multimodal, cross-model ensemble, and human-in-the-loop systems. Under specific benchmarks and controlled experimental settings, some studies report that multimodal detection frameworks integrating text, metadata, and contextual signals yield accuracy gains on the order of 5--10\% over text-only baselines \cite{liu2025modality}. Related work further explores ensemble-based verification strategies, reporting additional improvements (e.g., 8--12\%) for certain classes of subtle misinformation, though these gains depend on datasets and evaluation protocols and are not directly comparable across studies. In privacy protection, differential privacy~\cite{li2022dp} and linguistic steganography~\cite{bao2024keep} provide both theoretical and practical advances. }
Detection systems are evolving from single-model approaches to multimodal, ensemble, and explainable frameworks. By fusing text, images, and social network structures, multimodal systems can significantly improve the detection of complex and adversarial misinformation. Ensemble methods (e.g., model stacking, expert systems) can reduce individual model bias and enhance robustness against adversarial content. At the same time, explainability mechanisms provide transparent rationales for moderation decisions, fostering trust and usability for both moderators and end-users~\cite{zhang2023biasx}. 
Privacy protection is shifting from passive anonymization to proactive defense and dynamic monitoring. Techniques such as differential privacy, knowledge distillation, and tag-based prediction reduce raw data exposure, while systems like P-BERT and ProPILE enable real-time detection and blocking of sensitive information. 
Federated learning enables privacy-preserving model training across distributed systems without sharing raw data, while linguistic steganography provides methods for secure information encoding in collaborative settings~\cite{yao2024federated, bao2024keep}.
These methods must balance protection strength, system overhead, and user experience, and their effectiveness against advanced attacks remains an open question.

\subsection{Implications}

\textbf{Technical Implications.}
The technical implications of LLM deployment in social media manifest in three interconnected dimensions: content verification capabilities, adversarial content challenges, and platform security vulnerabilities. In terms of content verification, LLMs have transformed how platforms detect and combat false information. For example, Staab et al. \cite{staab2023beyond} and Collier et al. \cite{Collier2024-ng} demonstrate how these systems can effectively identify coordinated misinformation campaigns by analyzing subtle patterns across posts and user behaviors. However, this enhanced detection capability comes with its own risks---for example, the same pattern recognition capabilities can be exploited to breach user privacy and reconstruct sensitive information from seemingly innocuous social media interactions \cite{Kholoosi2024-uj}.
\textit{The second} dimension concerns the evolving nature of social media content manipulation. LLM-enhanced platforms face increasingly sophisticated forms of synthetic content and automated manipulation. As Duncan et al. \cite{Duncan2023-xk} suggests, these platforms must balance aggressive content filtering with the risk of false positives that could suppress legitimate discourse. This challenge is particularly apparent in cross-cultural contexts, where content moderation systems must adapt to diverse cultural norms while maintaining consistent integrity standards \cite{Lin2024-xa}.
\textit{The third} dimension addresses platform security and scalability. While federated approaches \cite{puppala2024socfedgpt} offer promising solutions for privacy-preserving content moderation, they introduce new vulnerabilities in cross-platform coordination. The emergence of sophisticated adversarial techniques targeting social media integrity systems\cite{nan2024let} necessitates continuous adaptation of defense mechanisms.

\textbf{Societal Implications.}
Based on our review, we also summarize three interconnected dimensions in terms of the societal impact, including platform trust dynamics, information equity, and collective behavior shifts. In the trust dimension, LLMs are fundamentally reshaping how users verify and consume social media content. While these systems enhance automated fact-checking capabilities, they also introduce new challenges to platform credibility. One typical example is that user trust can fluctuate dramatically following high-profile misinformation incidents, particularly when LLM-based detection systems fail to catch sophisticated synthetic content \cite{Heaton2024-aw}. This trust volatility is further complicated by cross-cultural variations \cite{Lian2024-ui}; that is, different societies exhibit varying levels of confidence in automated content verification systems, which further affect the overall effectiveness of platform integrity measures.

\color{black}
The second dimension concerns information equity and accessibility. While LLM-powered content verification promises more democratic access to fact-checking tools, their deployment often creates new forms of information disparities. Specifically, the effectiveness of these systems varies significantly across languages and cultural contexts \cite{Duncan2023-xk, Lin2024-xa}, potentially marginalizing users from non-dominant linguistic and cultural backgrounds. Such uneven performance can undermine community-level trust in automated verification systems and exacerbate perceptions of exclusion among affected populations. For instance, fact-checking systems may perform well on English-language misinformation but fail to detect similar content in Swahili or Tagalog due to limited training data or weaker language modeling support, raising broader concerns about inclusion and the societal legitimacy of AI-mediated information governance. 

\color{black}
{The third} dimension addresses collective behavior adaptation in response to LLM-enhanced content moderation. Social media users are developing new strategies for information sharing and verification, shaped by their interactions with automated content analysis systems. This has been widely demonstrated in prior studies \cite{Tlili2023-tn, Kertati2023-fk}, in which researchers have observed how communities adapt their communication patterns to either work with or circumvent LLM-based content filters, and reported the emergence of new forms of collaborative fact-checking that combine human expertise with LLM capabilities. 

\subsection{Future Research}

\subsubsection{Information Disorder Detection}

The future of information disorder detection and mitigation requires a comprehensive approach integrating technical innovations and adaptive systems.

\textbf{Technical Foundations and Detection Systems}
The cornerstone of future information disorder mitigation is the development of robust detection frameworks that fuse multiple data modalities and verification techniques. \textcolor{black}{Multi-modal detection systems that integrate text, metadata, and contextual signals are an active area of investigation. Under specific benchmarks and controlled experimental settings, some studies report accuracy gains on the order of 5--10\% over text-only baselines \cite{liu2025modality}. Related work explores cross-model verification via ensemble approaches that combine multiple LLMs, reporting additional improvements (e.g., 8--12\%) for certain classes of subtle misinformation, though these results depend on datasets and evaluation protocols and are not directly comparable across studies \cite{nature2024ensemble}.}These architectures are further enhanced by data synthesis and augmentation techniques, exemplified by FACT-GPT~\cite{choi2024automated} that improve detection capabilities through automated fact-checking and content verification.
As adversarial tactics evolve, robust defense mechanisms become crucial. Recent advances in adversarial training and prompt-level hardening, such as Prompt Adversarial Tuning~\cite{mo2024fight}, demonstrate promising approaches to enhancing model robustness while maintaining legitimate task performance. These developments can be coupled with improved prompt engineering and detection mechanisms~\cite{fi16080286,app14083532,li2024staying} to address evolving misinformation tactics effectively.

\textcolor{black}{\textbf{Co-evolutionary dynamics and governance.} The interaction between attackers and defenders can be modeled as a co-evolutionary process: attackers continuously adapt their generation and evasion strategies (e.g., prompt engineering, adversarial fine-tuning), while defenders improve detection, attribution, and policy intervention measures. Viewing this process as a resource-constrained game helps identify asymmetries (e.g., scale costs, latency constraints, false positive risks) and points to potential stable strategies. For example, increasing the marginal cost of large-scale abuse (e.g., rate limiting, attribution authentication), providing detection subsidies for smaller platforms, or implementing transparency requirements that increase the risk of attackers being detected. We suggest that future research should include: (a) formalizing attacker/defender payoffs, (b) simulating the arms race dynamics under different policy interventions, and (c) evaluating equitable governance mechanisms that can reduce systemic harm.}

\textcolor{black}{\textbf{Political economy of information disorder.} 
Technical solutions alone cannot resolve LLM information disorder without addressing the structural incentives of the ecosystem. Engagement-driven platform business models (attention, retention, advertising) can systematically reward sensational, emotionally manipulative, and polarizing content—precisely the type of content that LLMs can now generate at scale and rapidly iterate through optimization and A/B testing. At the same time, defensive capacity is economically uneven: high-quality moderation and detection require compute, model access, and sustained human review, which can advantage large platforms while leaving smaller platforms initiatives with higher relative costs and weaker safety coverage. These political economic asymmetries imply that best technical defenses may not be deployed where they are most needed, and that effective governance should be evaluated not only by accuracy but also by how it reshapes incentives and access.}

\textbf{Adaptive Learning and Cultural Context.}
Future systems must emphasize adaptability through continuous learning mechanisms that maintain accuracy on known patterns while quickly adapting to emerging threats. Meta-learning and parameter-efficient tuning approaches have shown potential in reducing annotation costs and accelerating responses to emergent narratives~\cite{yue2023metaadapt}. Integration of human feedback loops, through expert validation and model updates, can also help refine model accuracy and prevent the misclassification of benign content~\cite{yan2024orchestrate}.
Cross-cultural and linguistic adaptation represents another critical frontier. Language-agnostic, culturally aware embeddings have shown promise in reducing the gap between high- and low-resource settings~\cite{miao2024enhancing}. By developing enhanced embeddings and culturally adaptive techniques~\cite{rathinapriya2024adaptive,10.1093/llc/fqac049}, detection systems can achieve effective detection across diverse linguistic and cultural environments.

\textbf{User Interface and Behavioral Analysis.}
The success of information disorder mitigation depends critically on effective user interaction and behavioral understanding. Future interfaces must clearly communicate detection confidence levels and provide transparent explanations for flagged content. The BiasX framework demonstrates the value of free-text explanations in content moderation, significantly aiding moderators in identifying subtle toxic content~\cite{zhang2023biasx}. Understanding user behavior patterns, particularly through silent-user and propagation-path modeling~\cite{nan2024let}, enables more effective intervention strategies. Simulation models analyzing propagation dynamics~\cite{ijcai2024p873,li2024large} further enhance our ability to develop proactive mitigation approaches.

\textcolor{black}{{\textbf{Ground truth under synthetic reality.}} 
Current integrity systems implicitly assume a stable external ground truth against which claims can be verified~\citet{nissenbaum2019contextual}. However, the proliferation of generated content challenges this assumption by lowering the marginal cost of producing multiple, contradictory yet plausible narratives at scale. In such settings, verification cannot rely solely on content plausibility or closed-loop model agreement: detectors trained on synthetic data risk reinforcing self-referential signals detached from external evidence. This motivates shifting from binary “true/false” framing toward integrity as an evidence construct, emphasizing provenance, source reliability, and verifiable supporting context. Future research should therefore study how to maintain shared epistemic reference points online when synthetic information saturates the environment, and how integrity interventions can prioritize traceable evidence, auditability, and calibrated uncertainty rather than assuming a universally ground truth.}

\textcolor{black}{{\textbf{Intrinsic vs. extrinsic hallucinations.}} 
Recent work distinguishes between \emph{intrinsic hallucinations}, where generated content directly contradicts the provided source or context, and \emph{extrinsic hallucinations}, where the model introduces unsupported information that cannot be verified from the given evidence. Intrinsic hallucinations reflect failures of faithfulness, whereas extrinsic hallucinations reflect failures of grounding. Current LLM-based fact-checking systems are generally more effective at detecting intrinsic inconsistencies than resolving extrinsic claims that require trustworthy external knowledge. Future research should therefore prioritize scalable methods for addressing extrinsic hallucinations through robust retrieval, provenance-aware evidence integration, and verifiable grounding beyond model-internal consistency.}

\textcolor{black}{{\textbf{Real-time, multimodal, and interactive disinformation.}} 
An emerging frontier of information disorder extends beyond static text and images toward real-time and immersive manipulation, including deepfake video, voice cloning, and AI-driven avatars in live streaming and social VR environments. These settings pose qualitatively different threats, where deception unfolds interactively, and personalized persuasion can be delivered on the fly through photorealistic agents or cloned voices of trusted figures. Current post hoc forensic detection paradigms are unlikely to suffice for such low latency, conversational contexts. Future research should prioritize new integrity mechanisms for real-time environments, including live authentication and provenance protocols, robust avatar and voice verification, proof-of-humanity signals, and safeguards that can operate during interaction rather than only after harmful content has spread.}

\textcolor{black}{{\textbf{Toward More Generalized Intelligence for Information Integrity.}} 
Most existing approaches to information disorder detection are grounded in Artificial Narrow Intelligence (ANI), focusing on task-specific classification or verification under constrained threat models. While effective in controlled settings, such systems often struggle with long-horizon reasoning, cross-platform narrative tracking, and adaptive coordination against evolving adversaries. Emerging Artificial General Intelligence (AGI)-inspired capabilities, including richer world modeling, abstraction across domains, and goal-aware reasoning, offer potential avenues to enhance future integrity systems. These advances could enable more holistic reasoning over social context, co-evolutionary attacker–defender dynamics, and long-term governance interventions, particularly in cross-lingual and low-resource settings. At the same time, increased generality raises challenges related to controllability and oversight. Future research should therefore explore incremental integration of AGI-inspired components while maintaining clear human oversight and accountability.}

\subsubsection{LLM-enhanced Social Bot Detection}

The evolution of LLM-enhanced bot detection necessitates a new generation of detection technologies that are real-time and privacy-conscious.

\textbf{Advanced Detection Architectures}
The next generation of bot detection systems demands hybrid architectures that leverage multiple detection strategies. Research demonstrates that pipelines merging  ``digital DNA'' temporal signatures with transformer encoders yield superior performance over traditional graph-only detectors~\cite{Chawla2023-le}. Federated variants of these systems show particular promise in preserving privacy while enabling cross-platform intelligence sharing~\cite{yang2023fedack}. The integration of federated learning approaches, as demonstrated by Puppala et al.~\cite{Puppala2024-mx}, represents a crucial advancement in enhancing detection capabilities while maintaining robust privacy standards.

\textbf{Cross-modal and Multilingual Capabilities}
As bot operations increasingly span multiple languages and media types, future research must prioritize comprehensive cross-lingual and multimodal detection capabilities. Evaluation frameworks such as ETS-MM are advancing joint text-audio-visual embedding analysis~\cite{li2025ets}, while building on foundational work in cross-lingual detection by Panda et al.~\cite{Panda2021-bu} and Ahmed et al.~\cite{Ahmed2023-gh}. These developments enable more effective understanding and detection of malicious content across linguistic and cultural boundaries. The comprehensive evaluation frameworks pioneered by Gu et al.~\cite{Gu2024-wv} provide essential benchmarks for assessing multimodal bot detection effectiveness. 

\textcolor{black}{{\textbf{Cross-lingual and Low-resource Robustness:}} 
Cross-lingual integrity systems should be evaluated with stratified error analysis that reports precision/recall/false-positive rates by linguistic subgroup and stress tests known brittleness factors, including (i) \emph{script} variation, (ii) \emph{morphological complexity} and tokenization effects, (iii) \emph{code-switching} and mixed language discourse common in global online speech, and (iv) \emph{domain drift} that can differentially degrade performance across languages. Where primary studies do not report language by language breakdowns, we treat multilinguality as a key evidence gap rather than a resolved capability, and we explicitly call for standardized reporting of subgroup metrics and robustness ablations. We emphasize that multilingual transfer is not uniform across languages and is often evaluated on English and Twitter/X-centric benchmarks; therefore we treat cross-lingual robustness as an evidence gap unless stratified results are reported. We also highlight the need for broader evidence beyond Twitter/X and from non-Anglophone verification ecosystems to improve external validity in multilingual contexts.}

\textbf{Real-time Monitoring and Response.}
The advancement of real-time monitoring capabilities represents a critical direction for future research. Streaming graph transformers have achieved significant breakthroughs in reducing moderation latency to sub-second levels, meeting crucial requirements for live platforms~\cite{zhang2025transtreaming}. Recent innovations by Ramamonjisoa et al.~\cite{Ramamonjisoa2024-oh} and Bonechi et al.~\cite{Bonechi2024-zi} in automated moderation systems establish new benchmarks for real-time monitoring and response capabilities. These systems must balance immediate detection with accuracy and resource efficiency.

\subsubsection{Privacy Preservation}


The future of privacy preservation in LLM-enhanced social media environments demands a multi-layered framework integrating three key components: 

\textbf{Advanced Privacy-Preserving Architectures.}
The foundation of future privacy protection lies in developing sophisticated architectural solutions that can safeguard user data while maintaining system functionality. Differential privacy approaches show particular promise, with recent work by \citet{coffey2024differential} demonstrating how differentially private stochastic gradient descent during fine-tuning can effectively limit model memorization while preserving utility. Federated learning variants, as explored by \citet{yao2024federated}, offer promising solutions for distributed privacy protection, though they must carefully balance privacy guarantees with system performance. These advances must be complemented by innovative anonymization strategies that can adapt to evolving threats and content sensitivity levels.

\textbf{Contextual Integrity and Inference Protection.}
A critical challenge in current LLM systems is maintaining contextual privacy across diverse interaction scenarios. \textcolor{black}{The \textsc{CONFAIDE} benchmark reveals concerning levels of sensitive context leakage, with up to 57\% of tested scenarios showing vulnerability \cite{mireshghallah2022quantifying}. Future research must focus on developing robust inference-time privacy guards that can dynamically adapt to different contexts while preventing unauthorized information extraction. The \textsc{CONFAIDE} framework \cite{mireshghallah2022quantifying} provides a foundation for standardizing privacy evaluations, but more sophisticated protection mechanisms are needed to address emerging threats while preserving model utility. }Linguistic steganography techniques \cite{bao2024keep} and advanced PII protection methods \cite{jang2024development} show promise in providing granular privacy controls across different linguistic.

\textbf{Real-time Monitoring and Compliance.}
As privacy threats evolve and regulatory requirements become more stringent, particularly with frameworks like the EU AI Act \cite{euaiact2024}, real-time privacy monitoring and compliance systems become crucial. Systems like ProPILE \cite{kim2024propile} demonstrate the potential for proactive privacy protection through continuous monitoring and intervention. Future research must focus on developing explainable systems that can provide clear documentation of privacy measures while adapting to emerging regulatory requirements. 

\color{black}
\subsection{Ethical and Regulatory Perspectives}
Ethical and regulatory frameworks provide crucial guidance for deploying large language models in social media environments. IEEE Ethically Aligned Design emphasizes transparency, accountability, and human oversight, while the EU AI Act classifies many AI-driven moderation and risk assessment systems as high-risk and mandates requirements for risk management, bias mitigation, and monitoring. UNESCO’s Recommendation on the Ethics of AI highlights inclusiveness, fairness, and respect for linguistic and cultural diversity. Together, these initiatives underscore the need to align technological advances in LLM-based systems with ethical principles and regulatory requirements.
\color{black}

\section{Conclusions}

\textcolor{black}{This comprehensive review examines the critical challenges and future directions in social media information integrity within the context of LLM applications. Our systematic analysis reveals four critical challenges that define this landscape: the escalating sophistication of information disorder, the emergence of LLM-enhanced social bots, persistent biases and uneven performance across languages and cultural contexts, and the growing complexity of privacy preservation and accountability in LLM-based systems. Through detailed examination of current approaches and limitations, we identified how LLMs simultaneously serve as powerful tools for detecting malicious content while potentially enabling more sophisticated forms of synthetic content generation.}

\textcolor{black}{Future developments must focus on several key areas, each corresponding to the research gaps identified in this survey. To address the growing sophistication of information disorder, research should advance multi-modal and adaptive detection systems, including ensemble-based approaches that integrate heterogeneous signals under evolving threat models. To counter LLM-enhanced social bots and coordinated inauthentic behavior, future work must develop cross-platform coordination and real-time monitoring capabilities that capture propagation dynamics and adversarial adaptation at scale. To mitigate bias, multilingual performance gaps, and low-resource disparities, culturally aware models, stratified evaluation, and robustness analysis across languages and regions are essential. Finally, to respond to privacy, transparency, and accountability challenges, robust privacy-preserving mechanisms and governance-aware deployment practices are required to address contextual privacy leakage and ensure responsible system operation.}

\textcolor{black}{For social media platforms, these challenges necessitate comprehensive strategies combining policy, technical, and user protection measures. Platforms should develop clear guidelines for LLM-generated content, implement robust detection and moderation systems, and establish transparent content labeling mechanisms. Additionally, user education and protection should be prioritized through enhanced verification tools and effective reporting systems, particularly for communities that are disproportionately affected by biased or opaque automated decisions.}

\textcolor{black}{Overall, the success of these efforts will depend on the integration of technical solutions with policy frameworks, while maintaining transparency and user trust across social media platforms. As LLM technology continues to evolve toward more generalized and adaptive capabilities, the approaches outlined in this review provide a foundation for systematically addressing existing research gaps and building more secure, privacy-respecting, and trustworthy social media environments.}


\color{black}
\bibliographystyle{ACM-Reference-Format}
\bibliography{main}

\clearpage
\appendix
\section{Supplementary Material: Extended Landscape Analysis}

\color{black}

\subsection{Extended Landscape Analysis Topic 1: Information Disorder}
How LLMs Enhance Detection and Prevention.
LLMs' potential in misinformation detection has evolved from foundational approaches to sophisticated methods. As shown in Table~\ref{tab:misinfo_techniques}, detection techniques can be broadly categorized based on their methodology, including early BERT-based models and more advanced fine-tuned, knowledge-based approaches. Early BERT-based methods demonstrated initial capabilities in misinformation detection tasks~\cite{al2019justdeep,jiang2020modeling,dulhanty2019taking}. For example, Dulhanty et al.~\cite{dulhanty2019taking} explores automated methods through position detection, while Jiang et al.~\cite{jiang2020modeling} develops BERT-based transfer-learning models.

Strong capability as the LLMs have, their performance on unseen data or circumstances, named as zero-shot setting, can be further improved \cite{Chen2023-jf}. Recent advances have produced several specialized detection strategies. Fine-tuned models like FACT-GPT~\cite{choi2024automated}, which leverages GPT-4 generated datasets for automated fact-checking, and FactLLaMA~\cite{cheung2023factllama}, which employs instruct-tuning for automated verification, represent significant progress in detection capabilities. Knowledge-based approaches have further enhanced detection systems, with ChatGPT being used to construct knowledge-based semantic structures~\cite{yang2023rumor}. Multimodal approaches have emerged through the integration of LLaVA and CLIP for knowledge transfer~\cite{lee2024train}, while LEMMA~\cite{xuan2024lemma} combines Large Vision Language Models with Chain-of-Thought reasoning for comprehensive detection. Recent frameworks like MUSE~\cite{zhou2024correcting} attempt to address this through automated fact-checking using online credible sources, generating queries, and providing explanations with credible source links. 

Advanced adaptive techniques have also been developed, such as MetaAdapt~\cite{yue2023metaadapt}, which employs meta-learning for domain-adaptive few-shot detection, demonstrating superior performance over existing baselines. Zero-shot detection methods~\cite{Chen2023-jf} have shown promise in identifying both human-written and LLM-generated misinformation, with performance varying based on content characteristics.
Detection performance is influenced by multiple factors, with content length being a key determinant—longer texts often yield higher accuracy due to richer contextual cues~\cite{Chen2023-jf}. To address domain shifts and imbalanced data distributions, adaptive methods such as MetaAdapt have been developed~\cite{yue2023metaadapt}. In parallel, researchers have explored advanced prompting strategies and multi-step verification workflows~\cite{yang2023rumor,lee2024train}, incorporating multimodal tools like CLIP and LLaVA to further enhance model robustness~\cite{lee2024train}. Despite these advances, significant challenges remain. Detection systems still struggle with subtle manipulations and short-form content, while accuracy varies based on text origin and model architecture. The rapid evolution of generative models necessitates continuous updates to detection mechanisms, as evidenced by systems like FACT-GPT~\cite{choi2024automated} and FactLLaMA~\cite{cheung2023factllama}.

Social media platforms have fundamentally transformed fake news propagation, enabling rapid and extensive reach with minimal resources~\cite{9206973,10.1145/3404835.3462871,app12178398,article_1466830}. This accelerated dissemination significantly erodes public trust in both social media platforms and institutions~\cite{9207406,zhou2022multimodalfakenewsdetection,jin2024veracity}. Real-world examples of fake news reveal how easily misinformation can be framed to appear credible, emphasizing the need for timely and accurate detection mechanisms. The challenge is particularly acute in low-resource settings and across linguistic barriers. For example, Rathinapriya et al.~\cite{rathinapriya2024adaptive} reveal substantial accuracy drops when detection systems trained on high-resource languages were applied to regional Indian languages, emphasizing the critical gap in cross-lingual detection capabilities. The speed and scale of social media sharing make traditional fact-checking methods increasingly inadequate, as false information can reach massive audiences before corrections can be implemented~\cite{liu2024detect}.

The increasing prevalence of multimedia content presents significant technical challenges~\cite{cmc.2024.046202,fi16080286}. Current detection frameworks face multiple limitations: they rely heavily on human annotation, limiting scalability~\cite{gielensgoodbye}; they struggle with cross-platform content verification; and they often fail to capture context-dependent nuances. Additionally, generative AI advancements enable the creation of increasingly sophisticated and credible false content~\cite{li2024staying}, making detection more complex. These technical challenges are particularly evident in real-time detection scenarios, where the need for immediate response conflicts with the requirement for accurate verification.
The societal impact of fake news manifests across multiple critical domains. In healthcare, misinformation during the COVID-19 pandemic led to public confusion, vaccine hesitancy, and inadequate crisis responses~\cite{chen2021transformer,ayoub2021combat,10.1093/llc/fqac049}. In politics, fake news serves as a powerful tool for manipulation and polarization~\cite{stipiuc2024romanian,XING2024102534}, undermining democratic processes and public discourse. In economics, it disrupts markets and consumer behavior through false financial information and manipulated market signals~\cite{10477989,app14083532}. These problems are particularly severe in regions with limited media literacy~\cite{9644290,rathinapriya2024adaptive,liu2024detect}, where both active and passive consumers~\cite{nan2024let} can be significantly influenced by false narratives.
Fake news exploits emotional and cultural elements to enhance its credibility and impact~\cite{cmc.2024.046202,XING2024102534,10.1145/3589335.3651972}. Its sophisticated adaptation to regional contexts and local sensitivities maximizes societal impact~\cite{rathinapriya2024adaptive,liu2024detect}, particularly through multimedia elements that increase persuasiveness~\cite{cmc.2024.046202,fi16080286}. The emotional manipulation often targets existing societal divisions, amplifying conflicts and polarization. These challenges are further intensified by social media's connectivity~\cite{9206973,10.1145/3404835.3462871,app12178398,article_1466830}, where echo chambers and algorithmic content promotion can rapidly spread misinformation within susceptible communities, significantly affecting public discourse and social cohesion.
\subsection{Extended Landscape Analysis Topic 2: Social Bot}
LLM-enhanced bots demonstrate significant potential in maintaining and enhancing information integrity on social media platforms. These advanced bots serve as powerful tools for automated content verification, real-time fact-checking, and proactive misinformation prevention through their sophisticated pattern recognition and contextual analysis capabilities \citet{Xu2021-pl, Panagiotou2021-uj}. Their multilingual abilities enable cross-language content verification \cite{Garcia-Diaz2022-nu, Milkova2023-gw}, while their advanced natural language processing capabilities allow for more accurate identification of suspicious patterns in both user behavior and content distribution \cite{Garcia-Silva2021-hb, Ghanem2023-ku}. The integration of LLMs has particularly enhanced bots' ability to understand nuanced context, engage in interactive fact-checking, and provide immediate feedback with educational resources about potential misinformation \cite{Haynes2021-ga, Shafee2024-zc}. Emerging applications suggest a promising future where LLM-enhanced bots can participate in collaborative verification systems, implement adaptive learning mechanisms for improved detection accuracy, and deploy context-aware intervention systems for maintaining the health and reliability of social media information ecosystems \cite{Xu2021-pl, puppala2024socfedgpt}.

The evolution of LLM-enhanced bots and their impact on information integrity has been a growing concern since 2019. Early research by Gupta et al. \cite{Gupta2019-vf} and Ghanem et al. \cite{Ghanem2020-kg} revealed how malicious bots were increasingly adopting transformer-based architectures to generate more convincing spam content on Twitter. While their detection models achieved accuracy rates above 98\% for traditional bot behaviors, these studies also pointed out an emerging challenge: as bots incorporated more sophisticated LLM capabilities, they became increasingly difficult to distinguish from human users. Recent studies by Li et al. \cite{Li2023-et} and Yang et al. \cite{Yang2024-tq} further revealed how these bots evolved to form large-scale coordinated networks, manipulating information across various domains, from health misinformation to financial markets \cite{Tian2024-rw}.
The COVID-19 pandemic marked a critical period in the evolution of malicious bot capabilities. Multiple research teams documented the spread of health-related misinformation. Sai et al. \cite{Sai2020-lj} and Kar et al. \cite{Kar2020-dm} demonstrated how LLM-enhanced bots effectively disseminated COVID-19 misinformation across multiple languages. This was further confirmed by Kim et al. \cite{Kim2022-eg} and Sharma et al. \cite{Sharma2022-zq}, who found these bots achieved high accuracy in mimicking legitimate health information sources. Pranto et al. \cite{Pranto2022-uo} specifically identified ten distinct topics in fake news, illustrating the sophisticated nature of cross-lingual misinformation campaigns.
Recent advances in multimodal capabilities have further complicated the threat landscape. Lyu et al. \cite{lyu2023gpt} show how GPT-4V-powered bots can effectively manipulate both textual and visual content. This capability was further explored by Mehta et al. \cite{Mehta2024-xc}, who demonstrated how these bots form sophisticated user communities to amplify their impact. The integration of advanced language models has enabled these bots to generate highly personalized phishing content, as documented by Haynes et al. \cite{Haynes2021-ga} and Atzori et al. \cite{Atzori2024-tx}, who showed how bots exploit vulnerabilities across multiple social networks. The cross-lingual capabilities of these bots present particular challenges. Studies by Shukla et al. \cite{Shukla2023-jx} and Ahmed et al. \cite{Ahmed2023-gh} revealed how LLM-enhanced bots effectively adapt their content across different languages and cultural contexts. This adaptability was further demonstrated by Harrag et al. \cite{Harrag2021-bd} in their analysis of Arabic language manipulation, achieving deception rates of up to 98\%.

The development of defense mechanisms against LLM-enhanced bots has evolved into a multi-layered approach, incorporating various technological and methodological innovations. Advanced Detection Architectures have emerged as a primary defense strategy. \textcolor{black}{Alshattnawi et al.~\cite{Alshattnawi2024-ga} report improved spam detection performance using deep neural networks with contextualized word embeddings. In their experimental setting, gains on the order of 10--15\% over selected traditional baselines are observed, though these results are tied to specific datasets and evaluation protocols.} This work was complemented by Sangher et al. \cite{Sangher2024-pc}, who integrated LSTM and BERT-based transformers to identify sophisticated cybercrime patterns. The effectiveness of these approaches was further validated by Shafee et al. \cite{Shafee2024-zc}, who evaluated various LLM chatbots for cybersecurity threat awareness.
Proactive prevention systems represent another crucial defensive layer. Huntsman et al. \cite{Huntsman2024-iw} proposed innovative approaches using LLMs and sheaf theory to detect textual inconsistencies, while Milner et al. \cite{Milner2023-jc} developed lightweight phishing detection algorithms specifically optimized for mobile devices. These systems have been enhanced by Puppala et al. \cite{Puppala2024-mx}, who introduced a Federated Learning-based GPT system that ensures privacy while maintaining bot detection effectiveness.
Cross-platform and Multilingual Defense strategies have become increasingly important. Liu et al. \cite{Liu2023-gr} developed early identification methods using topic expansion and SBERT models, while Askari et al. \cite{Askari2024-cu} conducted field experiments to evaluate the effectiveness of bot-based interventions in promoting legitimate news consumption. These approaches demonstrate the importance of comprehensive, platform-agnostic defense mechanisms.

\textcolor{black}{To move beyond static veracity checks, we distinguish between three hierarchical detection tiers: (i) \textit{Account-level classification}, which focuses on profile metadata and historical longevity; (ii) \textit{Content-level detection}, leveraging LLMs to identify stylistic "fingerprints" and semantic inconsistencies; and (iii) \textit{Campaign-level discovery}, which identifies coordinated inauthentic behavior (CIB) through network-wide signals.}
\textcolor{black}{Standard bot detection benchmarks often suffer from inflated performance due to random sampling. We argue for more rigorous {negative sampling strategies}, such as selecting "active-benign" users who share high-frequency posting habits with bots but lack coordination signatures. Furthermore, to ensure robustness against {covariate shift}, models must be evaluated via {temporal splits} (e.g., training on historical election data and testing on current discourse) and {cross-platform transfer} (e.g., training on X/Twitter and testing on Mastodon). This reveals the "generalization gap" where detectors often fail as bot personas evolve.}
\textcolor{black}{While LLMs excel at semantic analysis, they are often blind to non-textual behavioral cues. A robust detection pipeline requires ablations on the following features:}
    {Network Homophily:} {Quantifying the tendency of bots to form dense, interconnected k-cores that are statistically distinct from human "small-world" networks.}
    {Temporal Burstiness:} {Measuring synchronized activity peaks across multiple accounts, which indicates automated scheduling or coordinated human-trolling.}
    {Action-Type Entropy:} {Humans typically exhibit high-entropy behavioral sequences (e.g., liking, then browsing, then replying), whereas bots often follow low-entropy, repetitive state transitions.}
\textcolor{black}{The "asymmetric co-evolution" of bot tactics leads to rapid {concept drift}. We identify {Test-Time Adaptation (TTA)} and {Continual Learning} as primary countermeasures. However, these introduce {pseudo-labeling risks}, where a model may inadvertently "self-train" on its own false positives, leading to the suppression of legitimate minority voices. Future research must prioritize uncertainty-aware models that can flag high-drift scenarios for human-in-the-loop review rather than relying on brittle, automated thresholding.}

\subsection{Extended Landscape Analysis Topic 3: Privacy}

Privacy-preserving techniques for LLM-enhanced environments show promising potential in addressing current privacy challenges. Knowledge distillation methods proposed by \citet{zhao2021knowledge} demonstrate how label prediction can effectively abstract content while minimizing sensitive data exposure. In URL filtering contexts, P-BERT introduced by \citet{supriya2022p} shows significant potential in integrating deep feature extraction with BERT for malicious link detection while reducing leakage risks. These advancements directly address the limitations identified by \citet{patsakis2023man} and \citet{mattern2022limits}, where traditional anonymization methods fail against LLMs' ability to infer user attributes from residual linguistic signals.

Advanced privacy protection mechanisms are emerging to tackle the issue of raw data exposure during processing. Differential privacy approaches show particular promise, with \citet{coffey2024differential} demonstrating how differentially private stochastic gradient descent during fine-tuning can effectively limit model memorization of individual data points. Privacy-preserving synthetic data generation, such as SynthPAI by \citet{yukhymenko2024synthetic}, offers a solution to avoid direct handling of personal information. Additionally, language-specific approaches for PII protection, as demonstrated by \citet{jang2024development} for Korean data, provide promising templates for extending privacy safeguards across different linguistic contexts.
Emerging solutions for deep anonymization are addressing the limitations of conventional de-identification methods. Linguistic steganography techniques \citet{coffey2024differential} show potential in encoding sensitive content in forms only intelligible to intended recipients, while authorship obfuscation frameworks \citet{bao2024keep} demonstrate how reinforcement learning can effectively conceal personal stylistic features without compromising meaning. These approaches directly address the challenges identified by \citet{patsakis2023man} and \citet{mattern2022limits} regarding LLMs' ability to infer user traits through subtle linguistic cues.
Real-time monitoring and accountability systems represent crucial advancements in privacy protection. For example, \citet{kim2024propile} and \citet{asimopoulos2024benchmarking} demonstrate potential in proactively detecting and suppressing sensitive information in model outputs. Federated learning approaches by \citet{yao2024federated, yao2019federated} show promise in reducing centralized data storage vulnerabilities, while explainable privacy frameworks proposed by \citet{mireshghallah2022quantifying} offer solutions for transparency and accountability concerns. These developments directly address the need for context-aware privacy reasoning in diverse and dynamic social media interactions, providing comprehensive solutions for protecting user data while maintaining system functionality.

In recent years, significant advances have been made in addressing LLM privacy challenges, with a focus on several prominent research directions. First, model inversion attacks \cite{morris2024lmi,carlini2021extracting,fredrikson2015model,wang2024promptinversion} exploit output probabilities to reconstruct hidden prompts or input data, posing serious threats in LLM-as-a-service settings where prompts may encode private instructions. In terms of data leakage risks, studies have revealed two major vulnerabilities: LLMs can retain and reproduce sensitive information even when not overfitted \cite{kandpal2023memorization,duan2025latent,satvaty2024undesirable,barbulescu2024each}, and through membership inference attacks, adversaries can determine whether specific user data was included in the training set using techniques like SPV-MIA with paraphrasing and self-calibrated reference models \cite{zhang2024mia,puerto2024scaling,chi2024shadow}. To address these challenges, researchers have developed various privacy-preserving approaches: differential privacy techniques \cite{li2022dp,huang2020instahide,wiseman2018learning} implement DP-SGD to provide formal privacy guarantees, while federated learning solutions \cite{su2024titanic} explore decentralized model adaptation to prevent centralized data exposure, though both approaches must balance privacy protection with model utility. These privacy vulnerabilities and protective measures represent the current landscape of privacy challenges in LLM-enhanced social media environments.

The integration of LLMs in social media applications presents critical privacy challenges that demand immediate attention. The primary challenge lies in preventing unauthorized information inference: \citet{staab2023beyond} reveal how LLMs can breach privacy by deducing sensitive personal information from pseudonymized content through subtle linguistic markers and reconstructing detailed user profiles by connecting seemingly unrelated pieces of information. The cross-platform privacy challenge is particularly pressing, as \citet{treves2023rurlman} exposes how tools like RURLMAN can breach user privacy by automatically linking identities across multiple platforms through shared URLs, creating comprehensive digital footprints without user consent. This vulnerability is further amplified by the challenge of preventing synthetic identity abuse, where \citet{ayoobi2023looming} identifies how these models enable the creation of deceptive profiles on professional networks like LinkedIn for malicious purposes. As \citet{dogan2023catch} emphasizes, there is a critical need to develop robust safeguards against the systematic linking of online personas with real-world identities, fundamentally challenging traditional privacy protection approaches.

Real-time deployment of LLMs introduces additional urgent privacy challenges that require innovative solutions. \citet{brown2024theory} demonstrates the pressing need to establish clear boundaries between private and public information in ChatGPT's live platform interactions, while \citet{dou2023reducing} underscores the challenge of preventing unauthorized behavior pattern analysis that could expose users' personal lives, health, and relationship information. In sensitive topic analysis, \citet{cai2023public} identifies the critical challenge of balancing analytical capabilities with privacy protection, particularly in mental health discussions where inadvertent exposure of personal insights poses significant risks. These challenges necessitate the immediate development of enhanced privacy frameworks, as \citet{martin2022ai} argues for stronger ethical guidelines in governing LLM deployment, and \citet{dickson2023ethics} emphasizes the urgent need to resolve the fundamental tension between analytical utility and privacy preservation in these sensitive contexts.

\end{document}